\documentclass[preprint2]{aastex}

\renewcommand*{\thefootnote}{\fnsymbol{footnote}}

\usepackage[multiple]{footmisc}
\usepackage{amsmath}
\usepackage{graphicx}

\shorttitle{PARSEC follow-up}
\shortauthors{Marocco et al.}

\begin{document}

\title{Parallaxes of Southern Extremely Cool objects (PARSEC)\\
II: Spectroscopic Follow-up and Parallaxes of 52 Targets\footnote{Based on observations obtained at the Southern Astrophysical Research (SOAR) telescope, which is a joint project of the Minist\'{e}rio da Ci\^{e}ncia, Tecnologia, e Inova\c{c}\~{a}o (MCTI) da Rep\'{u}blica Federativa do Brasil, the U.S. National Optical Astronomy Observatory (NOAO), the University of North Carolina at Chapel Hill (UNC), and Michigan State University (MSU).}}

\author{F. Marocco\altaffilmark{1}, A. H. Andrei\altaffilmark{1,2,3,4}, R. L. Smart\altaffilmark{3}, H. R. A. Jones\altaffilmark{1}, D. J. Pinfield\altaffilmark{1}, A. C. Day-Jones\altaffilmark{5}, J. R. A. Clarke\altaffilmark{1}, A. Sozzetti\altaffilmark{3},  P. W. Lucas\altaffilmark{1}, B. Bucciarelli\altaffilmark{3}, and J. L. Penna\altaffilmark{2}.
\altaffiltext{1}{Centre for Astrophysics Research, Science and Technology Research Institute, University of Hertfordshire, Hatfield AL10 9AB}
\altaffiltext{2}{Observat\'{o}rio Nacional/MCT, R. Gal. Jos\'{e} Cristino 77, CEP20921-400, RJ, Brasil}
\altaffiltext{3}{INAF/Osservatorio Astrofisico di Torino, Strada Osservatorio 20, 10025 Pino Torinese, Italy}
\altaffiltext{4}{Observat\'{o}rio do Valongo/UFRJ, Ladeira Pedro Ant\^{o}nio 43, CEP20080-090, RJ, Brasil}
\altaffiltext{5}{Departamento de Astronomia, Universidad de Chile, Camino del Observatorio 1515, Santiago, Chile}
}

\begin{abstract}
We present near-infrared spectroscopy for 52 ultracool dwarfs, including two newly discovered late-M dwarfs, one new late-M subdwarf candidate, three new L and four new T dwarfs. We also present parallaxes and proper motions for 21 of them. Four of the targets presented here have previous parallax measurements, while all the others are new values. This allow us to populate further the spectral sequence at early types (L0-L4). Combining the astrometric parameters with the new near-infrared spectroscopy presented here, we are able to investigate further the nature of some of the objects. In particular, we find that the peculiar blue L1 dwarf SDSS~J133148.92$-$011651.4 is a metal-poor object, likely a member of the galactic thick disk. We discover a new M subdwarf candidate, 2MASS~J20115649$-$6201127. We confirm the low-gravity nature of EROS-MP~J0032$-$4405, DENIS-P~J035726.9$-$441730, and 2MASS~J22134491$-$2136079. We present two new metal-poor dwarfs: the L4pec 2MASS~J19285196$-$4356256 and the M7pec SIPS2346$-$5928. We also determine the effective temperature and bolometric luminosity of the 21 targets with astrometric measurements, and we obtain a new polynomial relation between effective temperature and near-infrared spectral type. The new fit suggests a flattening of the sequence at the transition between M and L spectral types. This could be an effect of dust formation, that causes a more rapid evolution of the spectral features as a function of the effective temperature.
\end{abstract}


\section{Introduction}
\renewcommand*{\thefootnote}{\arabic{footnote}}
Among the challenges that modern astrophysics has to face, one of the most intriguing is the comprehension and modelling of the atmospheres of brown dwarfs. Discovered in large numbers by the deep optical and infrared surveys (DENIS, \citealt{1999A&A...349..236E}; SDSS, \citealt{2000AJ....120.1579Y}; 2MASS, \citealt{2006AJ....131.1163S}; UKIDSS, \citealt{2007MNRAS.379.1599L}; WISE, \citealt{2010AJ....140.1868W}), these extremely cool objects led to the extension of the spectral sequence, to include three new spectral types, L and T \citep{2005ARA&A..43..195K}, and Y \citep{2011ApJ...743...50C}. L dwarfs occupy the 2400-1400 K temperature range, and are characterized by extremely red colours, due mainly to the presence of dust in their atmospheres. T dwarfs are even cooler and, because their atmospheres are essentially free of dust that settles beneath the photosphere, they are bluer than the L dwarfs and their spectra are characterized by strong methane and water vapour bands. Y dwarfs are the coolest brown dwarfs known, and their spectra show almost equal flux in the J and H band and hints of NH$_3$ absorption in the blue wing of the H band peak.

The depletion of photospheric condensate clouds at the transition between the spectral classes L and T is one of the outstanding problems in brown dwarfs physics. In particular, current models are unable to explain the extremely narrow range of effective temperatures and luminosities in which this transition takes place \citep[e.g.][]{2006ApJ...640.1063B,2007ApJ...655..541M}. Also, our understanding of the effects of gravity and metallicity on the spectra of the cool dwarfs is still incomplete \citep[e.g.][]{2011MNRAS.414..575M,2012ApJ...748...74L,2012MNRAS.422.1922P}.

In order to examine the role of binarity, metallicity and gravity in the L-T transition region of the H.-R. diagram, it is necessary to combine spectroscopy, photometry and astrometry of a large sample of objects. For instance, binary candidates can be identified using spectral indices, and the spectral type of their components can be determined by spectral fitting. In particular, for objects in the L-T transition region, we refer the reader to \citet{2010ApJ...710.1142B}, where the authors developed a set of selection criteria based on a combination of spectral indices and spectral types. Unresolved binaries deserve particular attention as they are extremely important ``benchmark objects'', which can lead to dynamical masses measurements \citep[if their components can be spatially resolved, e.g.][and references therein]{2011ApJ...733..122D} or to radii measurements \citep[if they form an eclipsing pair, e.g.][]{2006Natur.440..311S}. Both quantities are required to put observational constraints on structure models and evolutionary theories of low-mass objects \citep{1998A&A...337..403B,2011ApJ...736...47B}. Metallicity and gravity can be estimated similarly by using spectral indices or via spectral fitting with benchmark objects \citep{2006MNRAS.368.1281P,2010ApJ...720L.113R}. Finally, a better sampling of the L and T spectral sequence is necessary to improve our understanding of the luminosity function and the substellar mass function, both still not well constrained \citep[e.g.][]{2010MNRAS.406.1885B}.

The PARallaxes of Southern Extremely Cool objects (PARSEC\footnotemark[1]\footnotetext[1]{http://parsec.oato.inaf.it}) program has been observing with the ESO2.2 Wide Field Imager (WFI) over 140 known L and T dwarfs to obtain their parallaxes and proper motions at a high S/N level. The observing campaign is complete and the project has already produced a proper motion catalogue of 220,000 objects and 10 parallaxes with 2 mas precision for the best cases \citep[hereafter AHA11]{2011AJ....141...54A}. A significant fraction of PARSEC targets do not have infrared spectroscopy, a lack that limits the depth of the analysis on the targets. We therefore started a spectroscopic campaign in parallel to the PARSEC program to follow-up those targets missing near-infrared (NIR) spectra.

In this contribution we present the first 52 spectra we obtained for PARSEC targets and new parallaxes and proper motions for 21 of them. The parameters of the sample can be found in Table \ref{list}, where we present objects' names, coordinates and infrared magnitudes. 

\begin{deluxetable}{l l c c c c c c c c c c} 
\tabletypesize{\scriptsize}
\setlength{\tabcolsep}{0.04in} 
\rotate
\tablewidth{0pt}
\tablecaption{List of the objects observed. \label{list}}
\tablehead{
Object name & Object & $\alpha$ & $\delta$ & 2MASS & 2MASS & 2MASS & WISE & WISE & WISE & WISE & Ref. \\
 & short name & hh:mm:ss.ss & dd:mm:ss.s & J & H & K$_s$ & W1 & W2 & W3 & W4 & }
\startdata
2MASS~J00145575$-$4844171   & 0014-4844 & 00:14:55.75 & -48:44:17.1 & 14.050 & 13.107 & 12.723 & 12.244 & 11.994 & 11.445 & 8.682 & 7 \\
EROS-MP~J0032$-$4405        & 0032-4405 & 00:32:55.84 & -44:05:05.8 & 14.776 & 13.857 & 13.269 & 12.820 & 12.490 & 11.726 & 9.289 & 1 \\
2MASS~J00531899$-$3631102   & 0053-3631 & 00:53:18.99 & -36:31:10.2 & 14.445 & 13.480 & 12.937 & 12.312 & 12.029 & 11.520 & 9.123 & 7 \\
2MASSW~J0058425$-$065123    & 0058-0651 & 00:58:42.53 & -06:51:23.9 & 14.311 & 13.444 & 12.904 & 12.562 & 12.248 & 11.692 & 8.739 & 2 \\
SSSPM~J0109$-$5100          & 0109-5100 & 01:09:01.50 & -51:00:49.4 & 12.228 & 11.538 & 11.092 & 10.833 & 10.573 & 10.373 & 9.309 & 3 \\
2MASS~J01282664$-$5545343   & 0128-5545 & 01:28:26.64 & -55:45:34.3 & 13.775 & 12.916 & 12.336 & 11.944 & 11.690 & 11.300 & 9.482 & 4 \\
2MASS~J01443536$-$0716142   & 0144-0716 & 01:44:35.36 & -07:16:14.2 & 14.191 & 13.008 & 12.268 & 11.603 & 11.361 & 10.948 & 8.928 & 18 \\
2MASS~J01473282$-$4954478   & 0147-4954 & 01:47:32.82 & -49:54:47.8 & 13.058 & 12.366 & 11.916 & 11.699 & 11.487 & 11.220 & 8.615 & 5 \\
2MASSI~J0218291$-$313322    & 0218-3133 & 02:18:29.13 & -31:33:23.0 & 14.728 & 13.808 & 13.154 & 12.599 & 12.287 & 11.926 & 9.415 & 6 \\
SSSPM~J0219$-$1939          & 0219-1939 & 02:19:28.07 & -19:38:41.6 & 14.110 & 13.339 & 12.910 & 12.546 & 12.307 & 12.868 & 9.176 & 3 \\
2MASS~J02271036$-$1624479   & 0227-1624 & 02:27:10.36 & -16:24:47.9 & 13.573 & 12.630 & 12.143 & 11.772 & 11.557 & 11.210 & 9.322 & 11 \\
2MASS~J02304498$-$0953050   & 0230-0953 & 02:30:44.98 & -09:53:05.0 & 14.818 & 13.912 & 13.403 & 12.943 & 12.700 & 11.901 & 9.481 & 5 \\
2MASSI~J0239424$-$173547    & 0239-1735 & 02:39:42.45 & -17:35:47.1 & 14.291 & 13.525 & 13.039 & 12.710 & 12.425 & 11.833 & 9.353 & 6 \\
2MASS~J02572581$-$3105523   & 0257-3105 & 02:57:25.81 & -31:05:52.3 & 14.672 & 13.518 & 12.876 & 12.018 & 11.591 & 10.596 & 8.952 & 7 \\
DENIS-P~J035726.9$-$441730  & 0357-4417 & 03:57:26.95 & -44:17:30.5 & 14.367 & 13.531 & 12.907 & 12.475 & 12.086 & 11.600 & 9.318 & 8 \\
SDSSp~J053951.99$-$005902.0 & 0539-0059 & 05:39:52.00 & -00:59:01.9 & 14.033 & 13.104 & 12.527 & 11.869 & 11.578 & 11.411 & 8.419 & 9 \\
SIPS0614$-$2019             & 0614-2019 & 06:14:11.96 & -20:19:18.1 & 14.783 & 13.901 & 13.375 & 13.044 & 12.789 & 12.563 & 9.344 & 10 \\
2MASS~J06244595$-$4521548	& 0624-4521 & 06:24:45.95 & -45:21:54.8 & 14.480 & 13.335 & 12.595 & 11.830 & 11.478 & 10.866 & 9.526 & 11 \\
2MASS~J07193535$-$5050523   & 0719-5050 & 07:19:35.35 & -50:50:52.4 & 10.327 &  9.735 &  9.482 &  9.270 &  9.122 &  9.012 & 8.609 & 12 \\ 
2MASS~J07193188$-$5051410   & 0719-5051 & 07:19:31.88 & -50:51:41.0 & 14.094 & 13.282 & 12.773 & 12.443 & 12.220 & 11.540 & 8.988 & 11 \\
SSSPM~J0829$-$1309			& 0829-1309 & 08:28:34.19 & -13:09:19.8 & 12.803 & 11.851 & 11.297 & 10.916 & 10.667 & 10.132 & 8.823 & 13 \\
2MASSW~J0832045$-$012835    & 0832-0128 & 08:32:04.15 & -01:28:35.8 & 14.128 & 13.318 & 12.712 & 12.411 & 12.173 & 11.809 & 9.141 & 2 \\
2MASSI~J0835425$-$081923    & 0835-0819 & 08:35:42.56 & -08:19:23.7 & 13.169 & 11.938 & 11.136 & 10.392 & 10.035 &  9.472 & 8.489 & 6 \\
DENIS-P~J0909$-$0658        & 0909-0658 & 09:09:57.49 & -06:58:18.6 & 13.890 & 13.090 & 12.539 & 12.207 & 11.957 & 11.300 & 8.572 & 19 \\
2MASSW~J0928397$-$160312    & 0928-1603 & 09:28:39.72 & -16:03:12.8 & 15.322 & 14.292 & 13.615 & 13.047 & 12.747 & 12.367 & 8.722 & 2 \\ 
2MASS~J09532126$-$1014205   & 0953-1014 & 09:53:21.26 & -10:14:20.5 & 13.469 & 12.644 & 12.142 & 11.757 & 11.404 & 10.761 & 8.719 & 14 \\
2MASS~J10044030$-$1318186   & 1004-1318 & 10:04:40.30 & -13:18:18.6 & 14.685 & 13.883 & 13.357 & 12.774 & 12.482 & 12.206 & 9.167 & 21 \\
2MASSW~J1004392$-$333518    & 1004-3335 & 10:04:39.29 & -33:35:18.9 & 14.480 & 13.490 & 12.924 & 12.285 & 11.998 & 12.667 & 9.223 & 15 \\
LHS~5166                    & LHS~5166  & 10:04:38.70 & -33:35:09.3 &  9.849 &  9.303 &  9.026 &  8.838 &  8.643 &  8.522 & 8.297 & 20 \\
2MASSI~J1045240$-$014957    & 1045-0149 & 10:45:24.00 & -01:49:57.6 & 13.160 & 12.352 & 11.780 & 11.452 & 11.227 & 10.762 & 8.957 & 15 \\
2MASSI~J1059513$-$211308    & 1059-2113 & 10:59:51.38 & -21:13:08.2 & 14.556 & 13.754 & 13.210 & 12.940 & 12.634 & 12.287 & 9.254 & 6 \\
2MASS~J11544223$-$3400390   & 1154-3400 & 11:54:42.23 & -34:00:39.0 & 14.195 & 13.331 & 12.851 & 12.350 & 12.037 & 11.369 & 9.548 & 8 \\
2MASS~J12462965$-$3139280   & 1246-3139 & 12:46:29.65 & -31:39:28.0 & 15.024 & 14.186 & 13.974 & 13.325 & 12.383 & 11.407 & 8.831 & 5 \\ 
SDSS~J133148.92$-$011651.4  & 1331-0116 & 13:31:48.94 & -01:16:50.0 & 15.459 & 14.475 & 14.073 & 13.412 & 13.123 & 12.262 & 9.481 & 16 \\
2MASS~J14044941$-$3159329   & 1404-3159 & 14:04:49.48 & -31:59:33.0 & 15.577 & 14.955 & 14.538 & 13.806 & 12.869 & 11.743 & 8.953 & 17 \\
2MASSW~J1438549$-$1309103   & 1438-1309 & 14:38:54.98 & -13:09:10.3 & 15.490 & 14.504 & 13.863 & 13.288 & 12.973 & 11.813 & 8.555 & 2 \\
SIPS1753$-$6559             & 1753-6559 & 17:53:45.18 & -65:59:55.9 & 14.095 & 13.108 & 12.424 & 11.837 & 11.519 & 11.127 & 9.383 & 10 \\
2MASS~J19285196$-$4356256   & 1928-4356 & 19:28:51.96 & -43:56:25.6 & 15.199 & 14.127 & 13.457 & 12.824 & 12.558 & 12.369 & 9.222 & 11 \\
2MASS~J19360187$-$5502322   & 1936-5502 & 19:36:01.87 & -55:02:32.2 & 14.486 & 13.628 & 13.046 & 12.278 & 11.998 & 11.646 & 8.146 & 11 \\
2MASS~J20025073$-$0521524   & 2002-0521 & 20:02:50.73 & -05:21:52.4 & 15.316 & 14.278 & 13.417 & 12.532 & 12.090 & 11.441 & 8.818 & 14 \\
2MASS~J20115649$-$6201127   & 2011-6201 & 20:11:56.49 & -62:01:12.7 & 15.566 & 15.099 & 14.572 & 14.431 & 14.117 & 12.371 & 9.196 & 5 \\
2MASS~J20232858$-$5946519   & 2023-5946 & 20:23:28.58 & -59:46:51.9 & 15.530 & 14.965 & 14.485 & 14.127 & 13.959 & 12.905 & 9.288 & 5 \\
SIPS2045$-$6332             & 2045-6332 & 20:45:02.38 & -63:32:06.6 & 12.619 & 11.807 & 11.207 & 10.738 & 10.358 &  9.860 & 8.682 & 10 \\
2MASS~J21015233$-$2944050   & 2101-2944 & 21:01:52.33 & -29:44:05.0 & 15.604 & 14.845 & 14.554 & 14.064 & 13.786 & 12.784 & 9.105 & 5 \\
2MASS~J21324898$-$1452544   & 2132-1452 & 21:32:48.98 & -14:52:54.4 & 15.714 & 15.382 & 15.268 & 14.955 & 13.635 & 12.014 & 8.733 & 5 \\
2MASS~J21481326$-$6323265   & 2148-6323 & 21:48:13.26 & -63:23:26.5 & 15.330 & 14.338 & 13.768 & 13.484 & 13.312 & 12.283 & 8.952 & 5 \\
2MASS~J21580457$-$1550098   & 2158-1550 & 21:58:04.57 & -15:50:09.8 & 15.040 & 13.867 & 13.185 & 12.571 & 12.226 & 11.656 & 8.472 & 7 \\
2MASS~J22092183$-$2711329   & 2209-2711 & 22:09:21.83 & -27:11:32.9 & 15.786 & 15.138 & 15.097 & 14.623 & 13.513 & 12.351 & 9.077 & 5 \\
2MASS~J22134491$-$2136079   & 2213-2136 & 22:13:44.91 & -21:36:07.9 & 15.376 & 14.404 & 13.756 & 13.229 & 12.832 & 11.552 & 9.070 & 14 \\
SSSPM~J2310$-$1759          & 2310-1759 & 23:10:18.46 & -17:59:09.0 & 14.376 & 13.578 & 12.969 & 12.593 & 12.285 & 12.106 & 8.693 & 3 \\
2MASS~J23185497$-$1301106   & 2318-1301 & 23:18:54.97 & -13:01:10.6 & 15.553 & 15.237 & 15.024 & 15.080 & 13.675 & 12.649 & 9.014 & 5 \\
SIPS2346$-$5928             & 2346-5928 & 23:46:26.56 & -59:28:42.6 & 14.515 & 13.905 & 13.500 & 13.252 & 12.925 & 12.279 & 9.182 & 10 \\
\enddata
\tablecomments{JHK magnitudes are from the 2MASS Point Source Catalogue.}
\tablerefs{(1) \citet{1999A&A...351L...5E}; (2) \citet{2000AJ....120..447K}; (3) \citet{2002A&A...389L..20L}; (4) \citet{2007MNRAS.374..445K}; (5) This paper; (6) \citet{2003AJ....126.2421C}; (7) \citet{2008ApJ...689.1295K}; (8) \citet{2003AJ....126.1526B}; (9) \citet{2000AJ....119..928F}; (10) \citet{2007A&A...468..163D}; (11) \citet{2008AJ....136.1290R}; (12) \citet{2007AJ....133.2898F}; (13) \citet{2002MNRAS.336L..49S}; (14) \citet{2007AJ....133..439C}; (15) \citet{2002ApJ...575..484G}; (16) \citet{2002AJ....123.3409H}; (17) \citet{2007AJ....134.1162L}; (18) \citet{2003AJ....125..343L}; (19) \citet{1999A&AS..135...41D}; (20) \citet{2002ApJS..141..187B}; (21) \citet{2010A&A...517A..53M}.} 
\end{deluxetable}

Ten of these targets were previously un-identified brown dwarfs (indicated as Ref. 5 in Table \ref{list}). They were selected as late-L and early-T candidates using 2MASS to provide near infrared colours, and combining this with Schmidt plate constraints from both USNO-B and the SuperCOSMOS Science Archive. We used the General Catalogue Query engine at the NASA/IPAC Infrared Science Archive to search the 2MASS database. Our 2MASS photometric constraints were designed to select ultracool objects over the range L8/9 to $\sim$T4. In general we selected 2MASS sources where J$\leqslant$16.0, 0.3$<$J-H$<$1.0, 0.0$<$H-K$<$0.9, 0.0$<$J-K$<$1.6, except for sources with the reddest J-H$>$0.8, where we instead imposed a limit of J$<$15.5. We required either non-detection in USNO-B or an R-band detection leading to a colour of R-K$>$8, with these constraints being implemented as part of our initial database search. In addition we excluded declinations of $<$-86 deg (since optical cross-matching in the database is incomplete in this range), and avoided the galactic plane by examining outside galactic latitudes between -15 and +15 deg. We also required no other 2MASS source within 6 arcseconds, no database evidence of contamination and confusion (cc$\_$flag=``000''), and no minor planet association (mp$\_$flg=``0''). This resulted in a large selection of candidates, dominated by contamination because our near-infrared colours constraints overlap greatly with stellar colours. The contamination took a variety of forms, including in the main part sources affected by bright star diffraction spikes, blended sources, and sources with faint (un-matched in the database) optical counterparts. To identify this contamination we visually inspected our full initial sample using the SuperCOSMOS Science Archive facility, and selected only candidates that were genuine non-detections in all bands, or if detected in the I-band, had colours consistent with late L or T dwarfs (I$-$J$>$3.5). Ten objects from this final selection form part of the sample investigated in this paper.

In Section 2 we present the astrometric results obtained for our sample. In Section 3 we describe the spectroscopic observing campaign, the strategy adopted and the reduction steps applied to the spectra, and we present the results obtained. In Section 4 we use the parallaxes and proper motion derived here to study the kinematics of our targets. In Section 5 we use spectra and parallaxes to determine the bolometric luminosity ($L_{\rm bol}$) and effective temperature ($T_{\rm eff}$) of the objects in the sample. In Section 6 we comment on the properties obtained for the individual objects. Finally in Section 7 we summarize the results obtained and we discuss the future analysis that we will perform on the sample.

\section{Astrometry}
The observing strategy adopted in PARSEC is described and discussed extensively in AHA11, and the reader is referred to that contribution for details.

The parallax solution also delivers the proper motion, based solely on the observations used for the parallax solution, thus reasonably independent from the previous result obtained by combining an early subset of these observations against the 2MASS position (AHA11). 

The objects in each image were centroided using the Cambridge Astronomy Survey Unit's \textit{imcore} maximum likelihood baricenter (CASUTOOLS, v 1.0.21).The ensuing astrometry is done in relative mode, that is selecting a reference frame and referring all others to this frame using the standard coordinates calculated from the measured centroids. In fact, we unbias the outcome from a priori choices by selecting every frame in turn as the reference frame, thus producing as many parallax solutions as frames. The parallax  and proper motion are calculated using the methods adopted in the Torino Observatory Parallax Program \citep{2003A&A...404..317S,2007A&A...464..787S} and identical to those in AHA11.

\begin{figure*}
\centerline{\includegraphics[width=8.5cm]{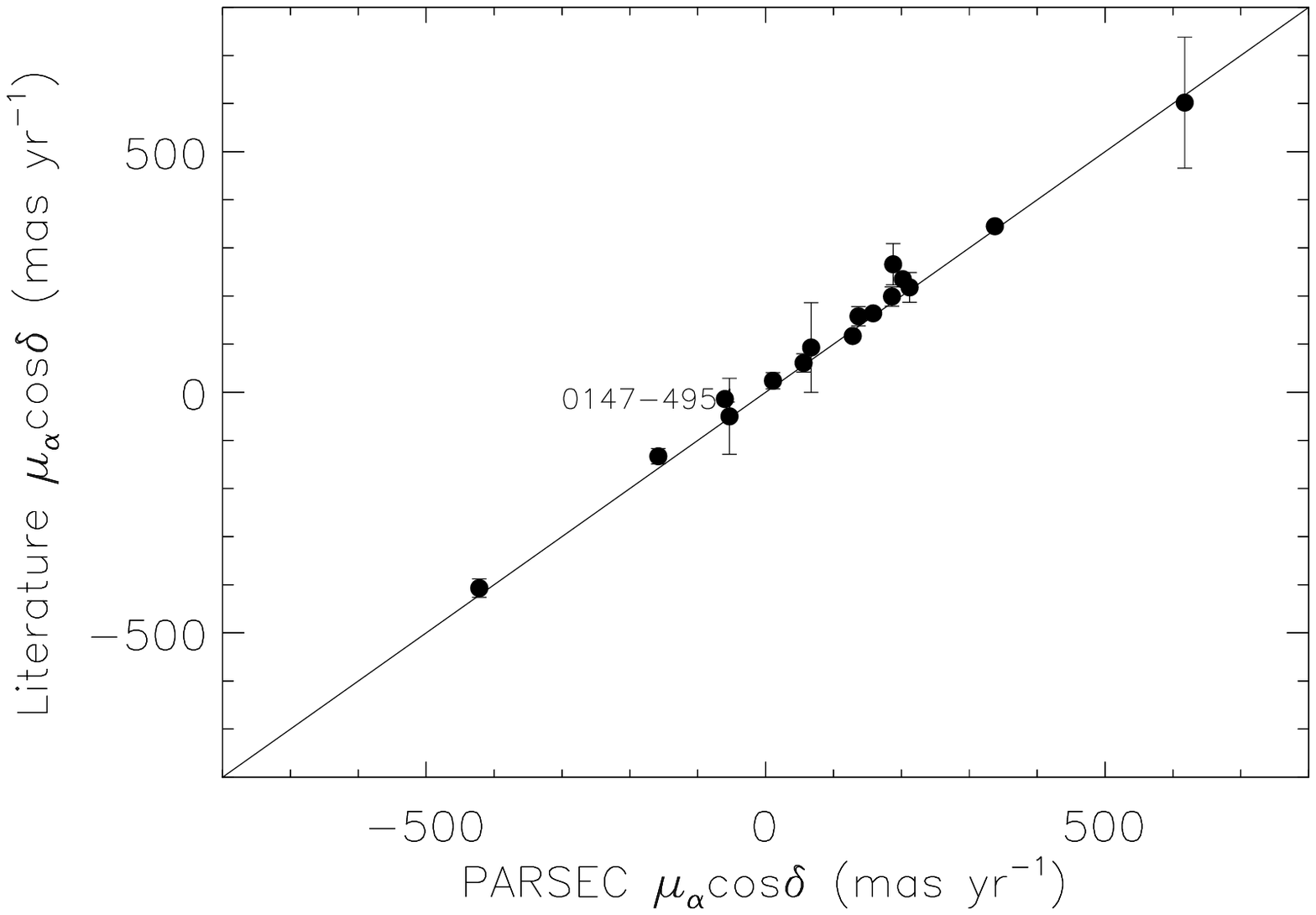}
\includegraphics[width=8.5cm]{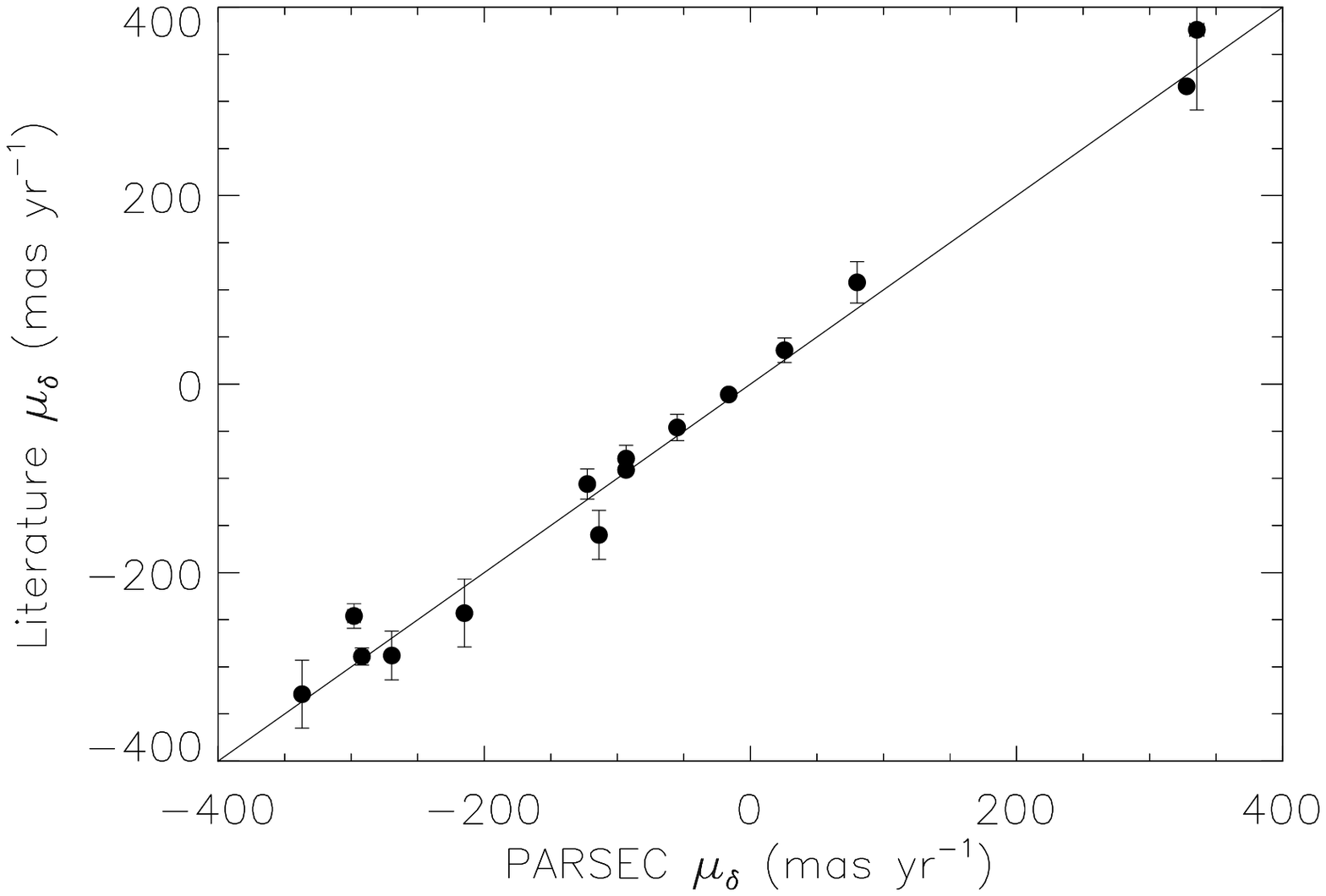}}
\caption{Comparison of proper motions obtained here with those published in the literature. The agreement is good, with only three targets (1404-3159, 1936-5502 and 0147-4954) showing a 3 $\sigma$ inconsistence in $\mu_\alpha$ and one (2310-1759) with a 3 $\sigma$ inconsistence in $\mu_\delta$. All the literature values are estimated from shorter baselines than those covered in this paper. \label{mu_parsec_vs_lit}}
\end{figure*}

We compared the proper motions to literature values as reproduced in Table \ref{mu_parsec_vs_lit}. Three targets, 1404-3159, 1936-5502 and 0147-4954, have differences in right ascension proper motion greater than 3 times the mean error and one target, 2310-1759, in declination proper motion. However, all the estimates are from short baselines of a few years and only one, 0147-4954, is outside 4 times the mean error so we believe this is reasonable consistency.  

We also found 4 objects with published parallaxes, all with short baseline programs. Of these only one, 1936-5502, differs by more than 3 times the mean error. The \citet{2012ApJ...752...56F} value is from only 1.31 years coverage which is the limit for disentangling proper motion and parallaxes and our experience is that often increased epoch coverage changes the value beyond the formal errors. We await the \citet{2012ApJ...752...56F} updated value before we consider this a significant difference.

\begin{figure*}
\centerline{\includegraphics[width=8.5cm]{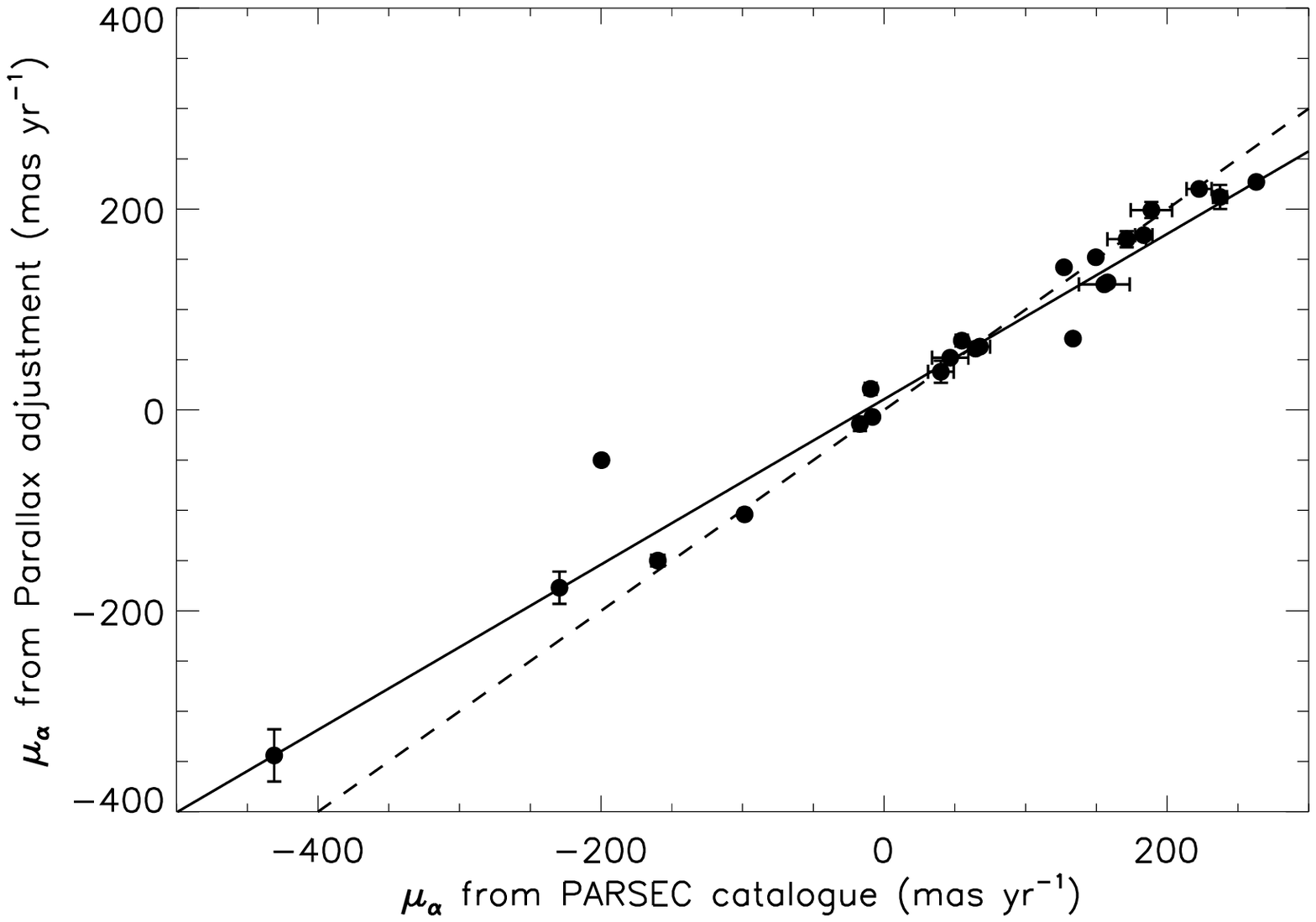}
\includegraphics[width=8.5cm]{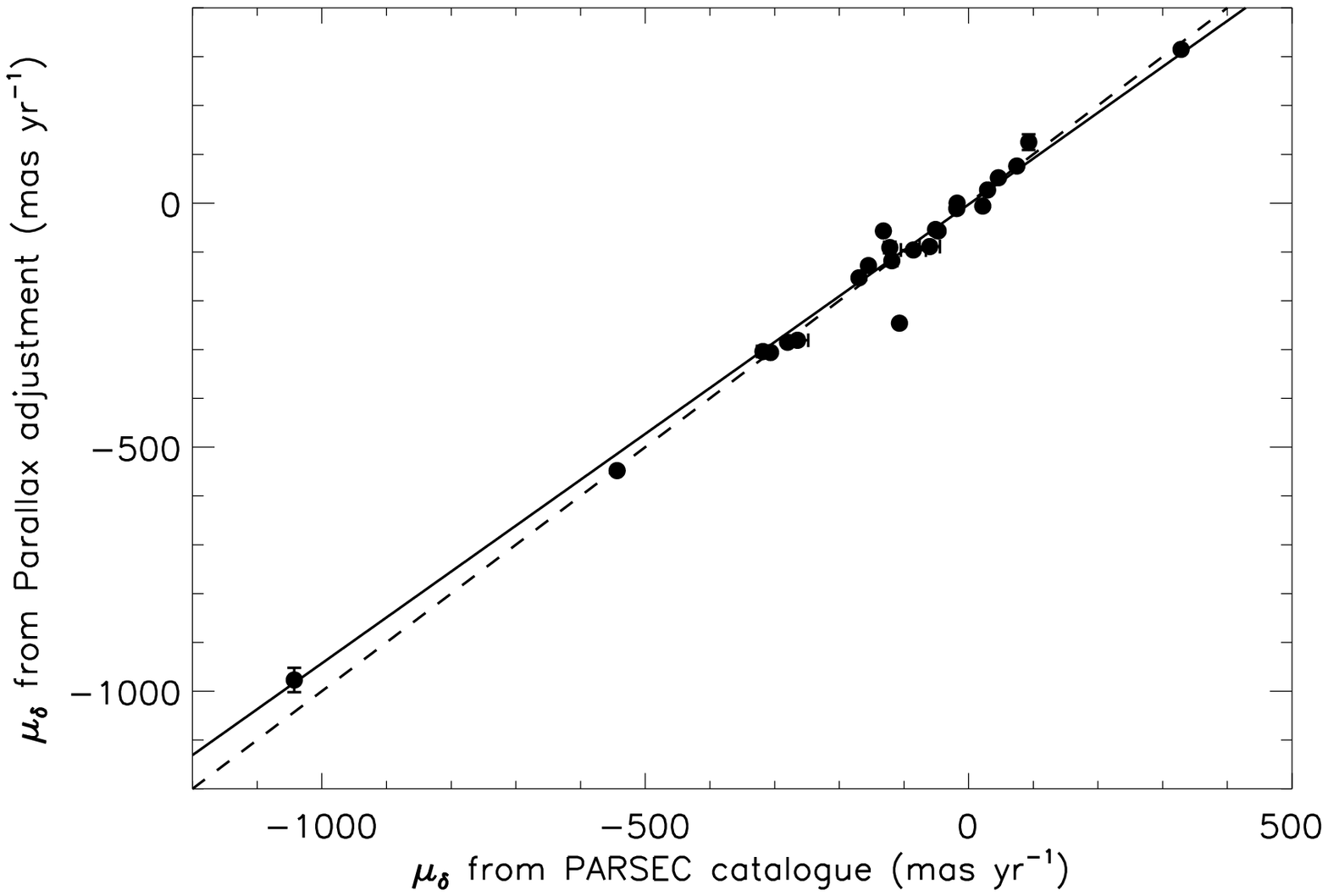}}
\caption{Comparison of the proper motions from the PARSEC published catalogue (AHA11) against the proper motions obtained in the parallax determination. Overplotted for reference are the bisector of the plot (dashed line) and a linear fit to the data (solid line). The angular coefficients of the linear fit are larger than 0.8 in both cases. \label{parallax1}}
\end{figure*}

The two panels of Figure \ref{parallax1} compare respectively the right ascension and declination proper motions obtained here against the values obtained in the PARSEC proper motion catalogue, which uses the subsample of the first 1.5 yr of PARSEC observations and the 2MASS positions, to a total time span of about 10 yr. It is clear that the agreement is good, with a linear fit of angular coefficient larger than 0.8. The significance of the Pearson correlation goes as $r*(\sqrt{N-2}/\sqrt{1-r^2})$ that is, for a given angular coefficient $r$ the larger the number of $N$ pairs the more significant is $r$. In our case with $N$ = 23, $r$ = 0.8 is significant to the  99.5$\%$ level. This lends support to the methods and significance to the assigned errors. Notice also that for 6 targets there was no corresponding proper motion in the PARSEC catalogue, meaning that they were either not found or not uniquely found in the 2MASS comparison. 

Proper motions and parallaxes of the targets are listed in Table \ref{astro}. For each target we present short name, the absolute parallax ($\pi_{abs}$), the two components of the proper motion ($\mu_{\alpha}$cos$\delta$ and $\mu_{\delta}$), the time span covered by the observations, and the number of observations available for each target ($N_{obs}$). 

\begin{deluxetable}{l c c c c c c}
\tablewidth{0pt}
\tablecaption{Summary of astrometric results. \label{astro}}
\tablehead{Object & $\pi_{abs}$ & $\mu_\alpha$cos$\delta$ & $\mu_\delta$ & $V_{\rm tan}$ & Time & $N_{obs}$ \\
short name & (mas) & (mas yr$^{-1}$) & (mas yr$^{-1}$) & km s$^{-1}$ & span (yr) & }
\startdata
0032-4405 & 21.6 $\pm$ 7.2 &  128.3 $\pm$ 3.4 &    -93.4 $\pm$ 3.0 & 34.8 $\pm$ 11.6 & 3.88 & 23 \\
0058-0651 & 33.8 $\pm$ 4.0 &  136.7 $\pm$ 2.0 &   -122.6 $\pm$ 1.8 & 25.8 $\pm$ 3.0  & 3.88 & 24 \\
0109-5100 & 57.8 $\pm$ 3.5 &  212.0 $\pm$ 1.7 &     80.2 $\pm$ 3.2 & 18.6 $\pm$ 1.1  & 3.88 & 25 \\
0147-4954 & 26.6 $\pm$ 3.1 &  -60.1 $\pm$ 1.9 &   -269.5 $\pm$ 1.8 & 49.2 $\pm$ 5.8  & 3.20 & 16 \\
0219-1939 & 37.2 $\pm$ 4.1 &  187.8 $\pm$ 2.5 &   -113.8 $\pm$ 3.3 & 27.9 $\pm$ 3.0  & 2.62 & 16 \\
0230-0953 & 32.4 $\pm$ 3.7 &  148.2 $\pm$ 1.9 &    -39.1 $\pm$ 2.7 & 22.4 $\pm$ 2.6  & 3.30 & 21 \\
0239-1735 & 32.1 $\pm$ 4.7 &   55.8 $\pm$ 2.2 &    -93.4 $\pm$ 2.2 & 16.1 $\pm$ 2.4  & 3.31 & 22 \\
0257-3105 & 99.7 $\pm$ 6.7 &  617.3 $\pm$ 3.6 &    335.5 $\pm$ 5.3 & 33.4 $\pm$ 2.2  & 3.09 & 13 \\
0539-0059 & 79.1 $\pm$ 2.4 &  158.3 $\pm$ 1.6 &    327.8 $\pm$ 2.4 & 21.8 $\pm$ 0.7  & 3.46 & 23 \\
0614-2019 & 34.3 $\pm$ 3.0 &  138.8 $\pm$ 2.0 &   -294.4 $\pm$ 2.9 & 45.0 $\pm$ 4.0  & 3.46 & 35 \\
0719-5051 & 34.6 $\pm$ 2.2 &  186.0 $\pm$ 1.2 &    -55.1 $\pm$ 1.7 & 26.6 $\pm$ 1.7  & 3.95 & 46 \\
0928-1603 & 34.4 $\pm$ 3.9 & -158.1 $\pm$ 2.1 &     25.6 $\pm$ 1.8 & 22.0 $\pm$ 2.5  & 3.94 & 23 \\
1246-3139 & 87.3 $\pm$ 3.2 &   -5.3 $\pm$ 1.7 &   -562.5 $\pm$ 2.6 & 30.5 $\pm$ 1.1  & 3.06 & 21 \\
1331-0116 & 67.3 $\pm$ 12.6 & -421.9 $\pm$ 5.7 & -1039.0 $\pm$ 5.2 & 79.0 $\pm$ 14.8 & 3.39 & 17 \\
1404-3159 & 49.2 $\pm$ 3.4 &  337.6 $\pm$ 1.9 &    -16.3 $\pm$ 2.4 & 32.5 $\pm$ 2.2  & 3.39 & 24 \\
1753-6559 & 58.0 $\pm$ 4.9 &  -53.3 $\pm$ 3.0 &   -336.9 $\pm$ 2.2 & 27.9 $\pm$ 2.4  & 4.28 & 55 \\
1936-5502 & 43.3 $\pm$ 4.5 &  202.0 $\pm$ 2.9 &   -292.0 $\pm$ 4.3 & 38.9 $\pm$ 4.1  & 3.88 & 40 \\
2045-6332 & 40.0 $\pm$ 3.7 &   67.0 $\pm$ 2.4 &   -214.9 $\pm$ 3.2 & 26.6 $\pm$ 2.5  & 3.87 & 25 \\
2209-2711 & 47.9 $\pm$ 12.5 &  -5.9 $\pm$ 8.1 &   -133.6 $\pm$ 9.9 & 13.2 $\pm$ 3.6  & 2.96 & 15 \\
2310-1759 & 36.4 $\pm$ 6.9 &   10.7 $\pm$ 5.4 &   -297.9 $\pm$ 4.7 & 38.8 $\pm$ 7.4  & 2.06 &  8 \\
2346-5928 & 14.3 $\pm$ 3.4 &  245.1 $\pm$ 1.7 &     57.6 $\pm$ 1.9 & 83.5 $\pm$ 19.7 & 3.88 & 24 \\
\enddata
\tablecomments{For each target we present short name, the absolute parallax ($\pi_{abs}$), the two components of the proper motion ($\mu_{\alpha}$cos$\delta$ and $\mu_{\delta}$), the time span covered by the observations, and the number of observations available for each target ($N_{obs}$).}
\end{deluxetable}

\section{Spectroscopy}

\subsection{Observations and Reduction Procedures}

Fourty-five of the spectra were obtained using the OSIRIS spectrograph on the SOAR telescope in low-resolution (R = 1200) cross-dispersed mode, covering the wavelength range 1.2-2.3 $\mu$m. The data were reduced following standard procedures. The spectra were flat-fielded using dome flats, dark subtracted, and pair-wise subtracted to remove sky lines. The extraction was performed using IRAF standard routines and the wavelength calibration was done with He-Ar arc lamps. In order to correct the measured spectra for the telluric absorption, standard stars were observed immediately before or after each target, close on the sky and at a similar airmass. The spectra were corrected dividing each of them by the spectrum of the associated standard and then multiplying the result by the theoretical SED from \citealt{1993IAUCB..21...93K} (for the appropriate temperature and surface gravity). The different orders of the telluric corrected spectra (roughly coincident with J, H and K band) were then merged, using the overlapping regions to adjust the relative flux levels, and finally turned into an absolute flux scale using the measured magnitudes (2MASS H and K$_{s}$). To do that, we convolved the spectra with 2MASS filters' profiles and integrated over the passbands to obtain synthetic magnitudes. Given that the difference between two magnitudes is, by definition, $m_{1}$-$m_{2}$=2.5$\times\log_{10}(f_{1}/f_{2})$, where $m_1$ and $m_2$ are the apparent magnitudes and $f_1$ and $f_2$ the corresponding fluxes, the scaling factors ($sf_{\textrm{i}}$) are given by the equation:

\begin{equation}
sf_{\textrm{i}} = 10^{0.4 \times \left( m_{i,synt}-m_{i,obs} \right) }
\end{equation}

where $m_{i,obs}$ is the measured 2MASS magnitude in the i$th$ band (H or K$_s$) and $m_{i,synt}$ is the corresponding synthetic one. We use H and K$_s$ band only, as the spectral coverage of OSIRIS is insufficient to compute a synthetic J magnitude. Finally, after checking that the two values were consistent, we took their weighted average as our scaling factor. To check the accuracy of our flux calibration, we tried to use different telluric stars taken during the same night (if possible) for each target. We noticed that there are no significant differences on the flux level outside of the telluric absorption bands. However, using a different telluric star significantly affects the quality of the telluric corrections, resulting in noisy telluric bands and significant variations in their flux level (up to a factor of 2).

Two other spectra were obtained with SOFI, on NTT, using a blue grism at low resolution (R = 1000) covering the wavelength range 0.95-1.64$\mu$m. The spectra reduction follows the same steps as for the OSIRIS ones, but the wavelength calibration was done using Xe arc lamps, and the flux calibration used the J magnitude only.

Finally five spectra were obtained with Xshooter, the echelle spectrograph mounted on the UT2 at VLT. This instrument covers a wide wavelength range (0.3-2.48 $\mu$m) with a resolution of 8100 in the VIS arm and 5500 in the NIR arm. To reduce these targets we used the Xshooter pipeline (version 1.3.7). The details of the Xshooter data reduction can be found in \citet{2013MNRAS.430.1171D}, and here we briefly summarize the main steps. The pipeline performs all the standard reduction steps (flat fielding, dark subtraction, wavelength calibration and flux calibration) and produces a 2D image containing the reduced spectrum. We extracted the spectra using standard IRAF routines and we corrected them for telluric absorption using standard telluric stars observed during the night, following the procedure described above. The telluric stars were also processed using the Xshooter pipeline. We tested the accuracy of the the telluric correction by using different telluric stars observed during the same night. As for the OSIRIS spectra, the use of different standards does not affect the global flux calibration, but results in some cases in slightly different flux levels in the telluric bands.

\begin{figure*}
\includegraphics[width=16cm,height=20cm]{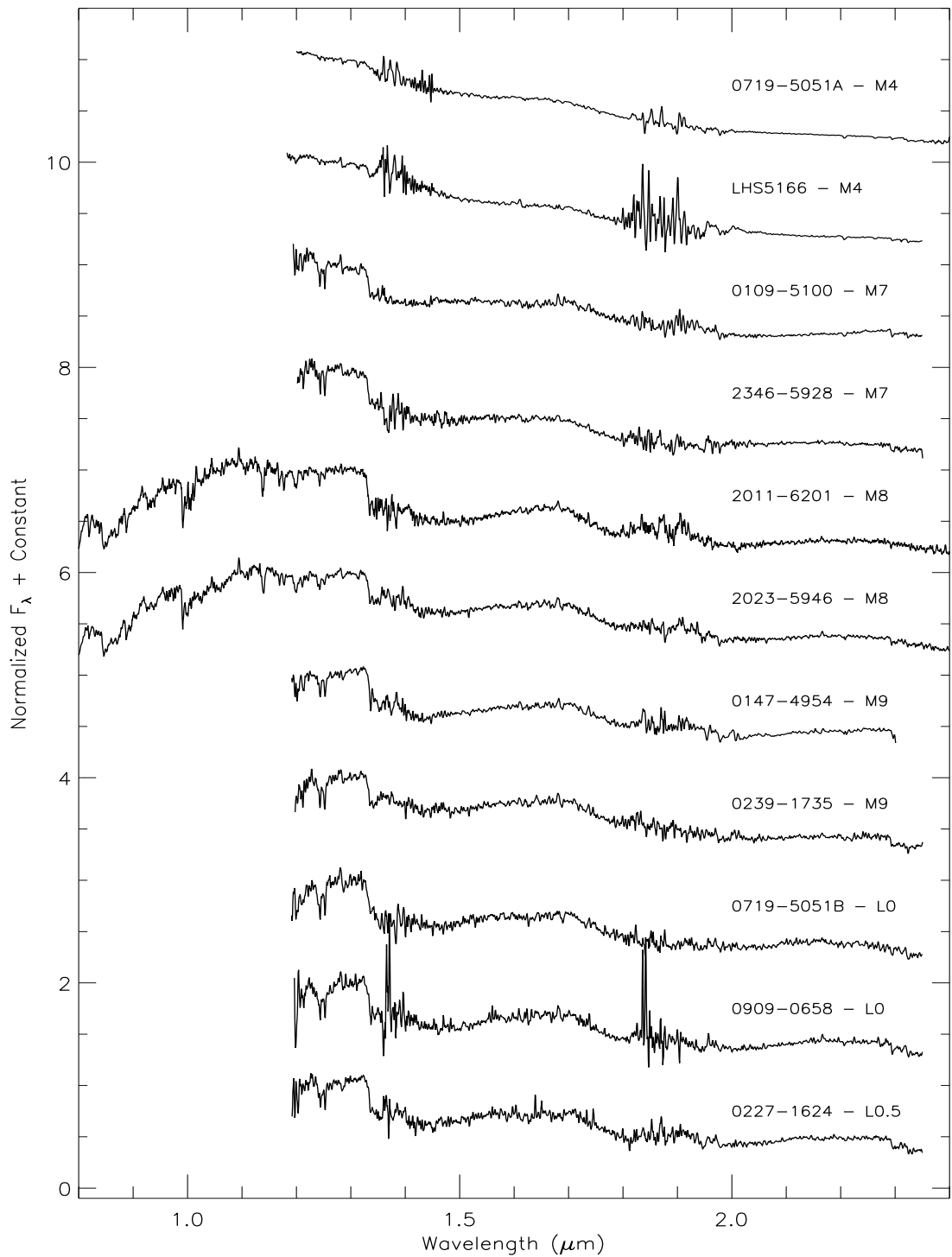}
\caption{The spectra obtained for our targets, sorted from earlier to later spectral type. The spectra showed here are in the M4-L0.5 range. They have all been normalized to 1 at 1.28$\mu$m, smoothed to a resolution of $\sim$10 \AA{} per pixel, and displaced vertically by increments of one flux unit. \label{spectra1}}
\end{figure*}

\begin{figure*}
\includegraphics[width=16cm,height=20cm]{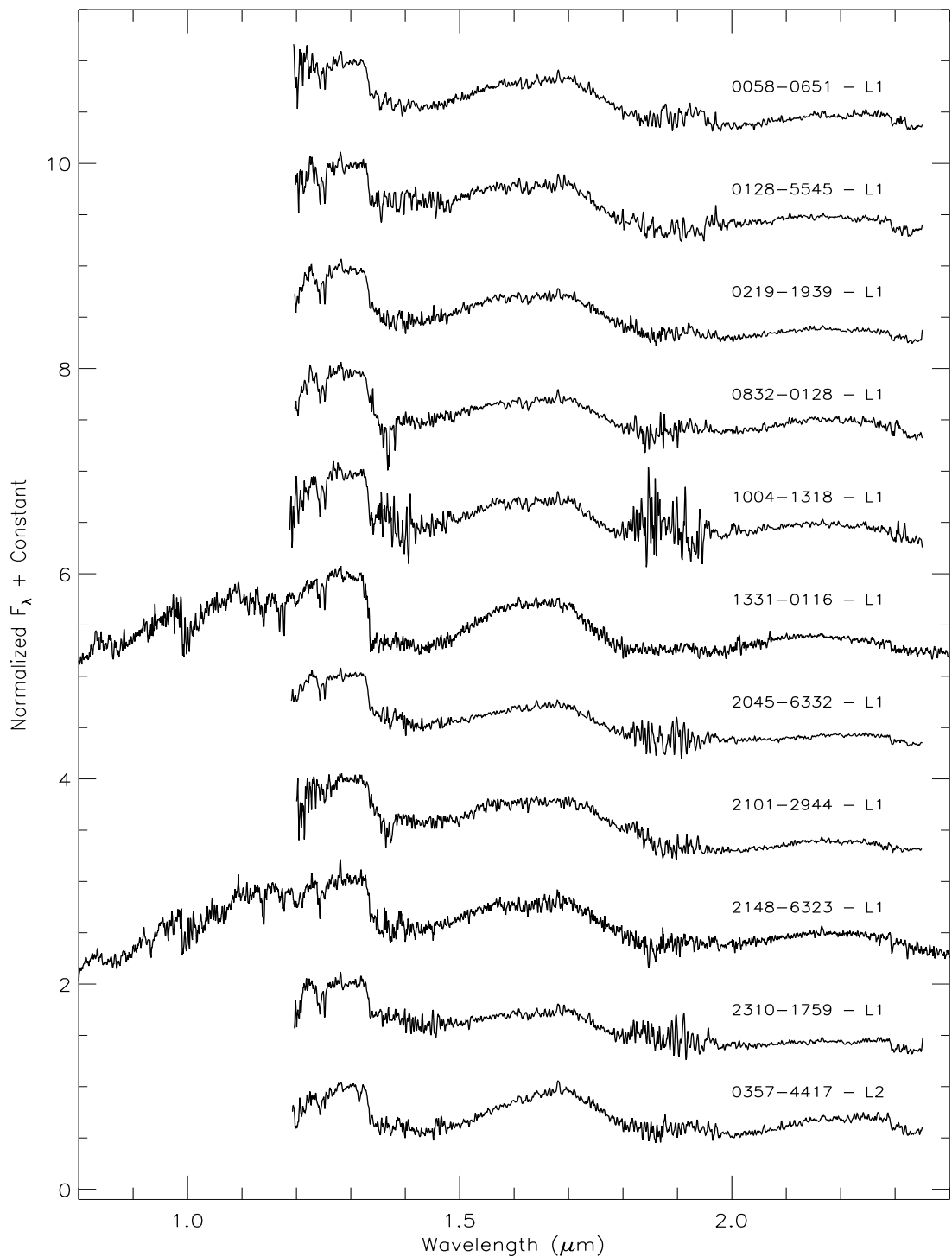}
\caption{Same as Figure \ref{spectra1}, but for objects in the range L1-L2. \label{spectra2}}
\end{figure*}

\begin{figure*}
\includegraphics[width=16cm,height=20cm]{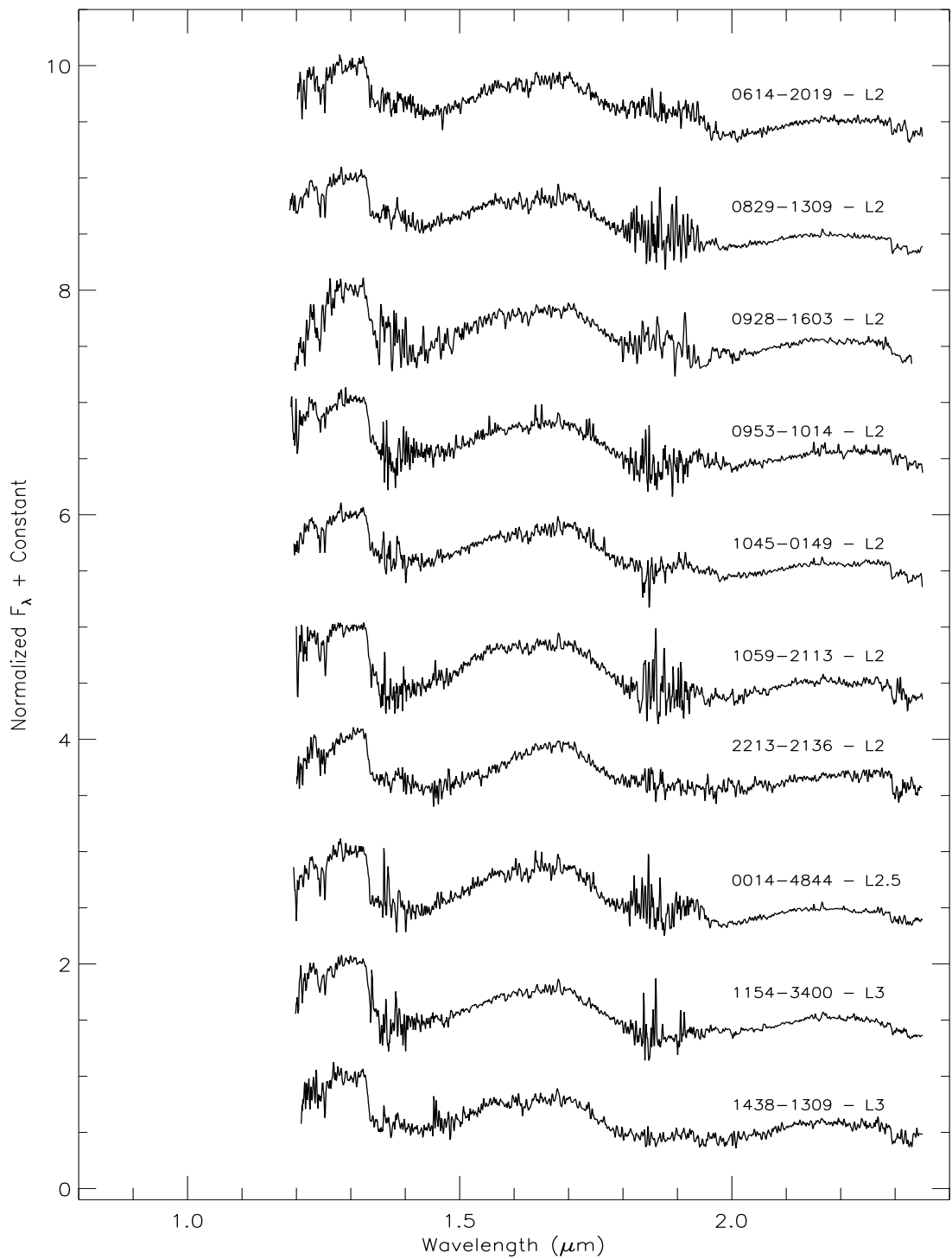}
\caption{Same as Figure \ref{spectra1}, but in the range L2-L3. \label{spectra3}}
\end{figure*}

The spectra obtained are presented in Figures \ref{spectra1} - \ref{spectra3}. All the spectra are normalized to 1 at 1.28 $\mu$m and shifted vertically by increments of one flux unit.

\begin{figure*}
\includegraphics[width=16cm,height=20cm]{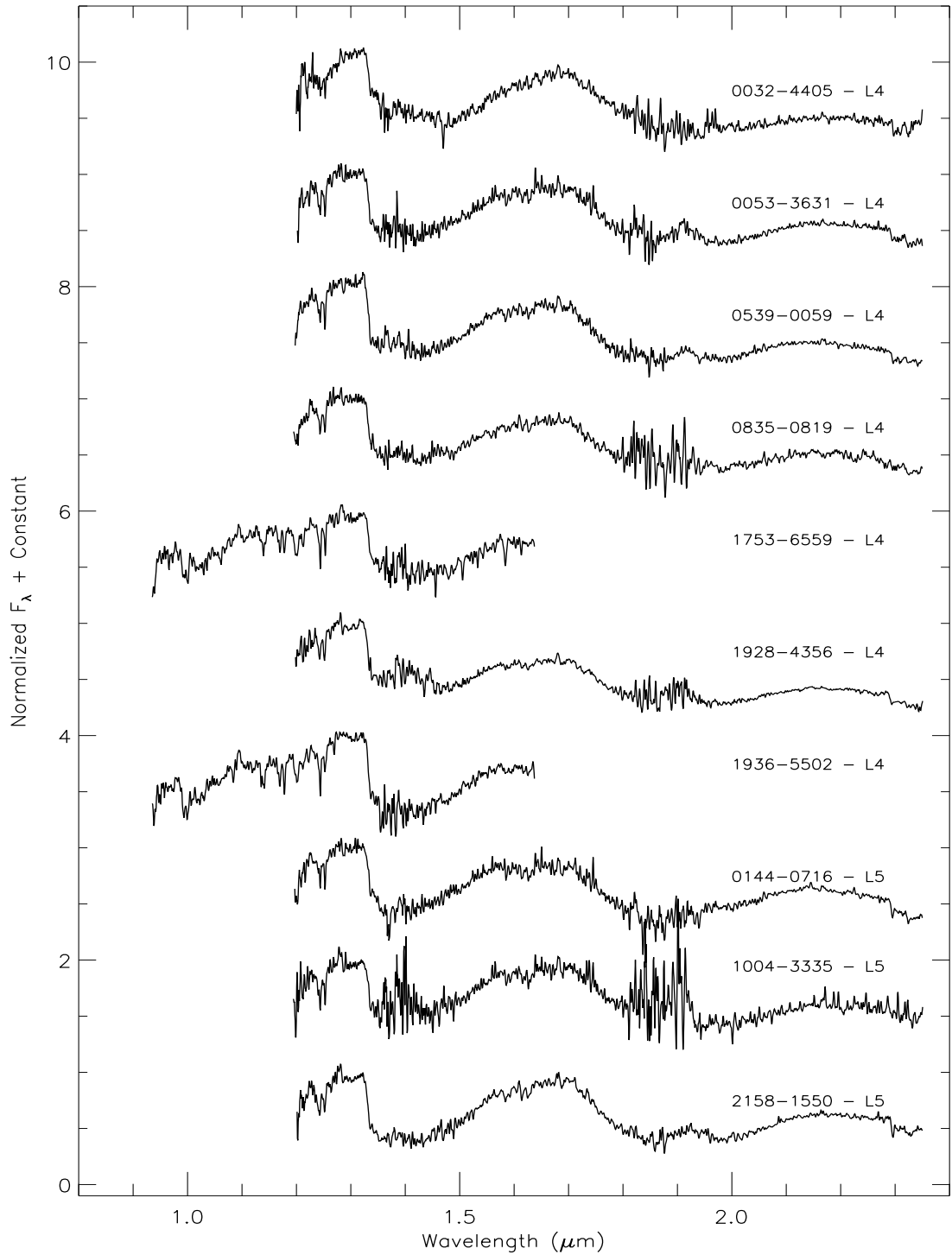}
\caption{Same as Figure \ref{spectra1}, but in the range L4-L5. \label{spectra4}}
\end{figure*}

\begin{figure*}
\includegraphics[width=16cm,height=20cm]{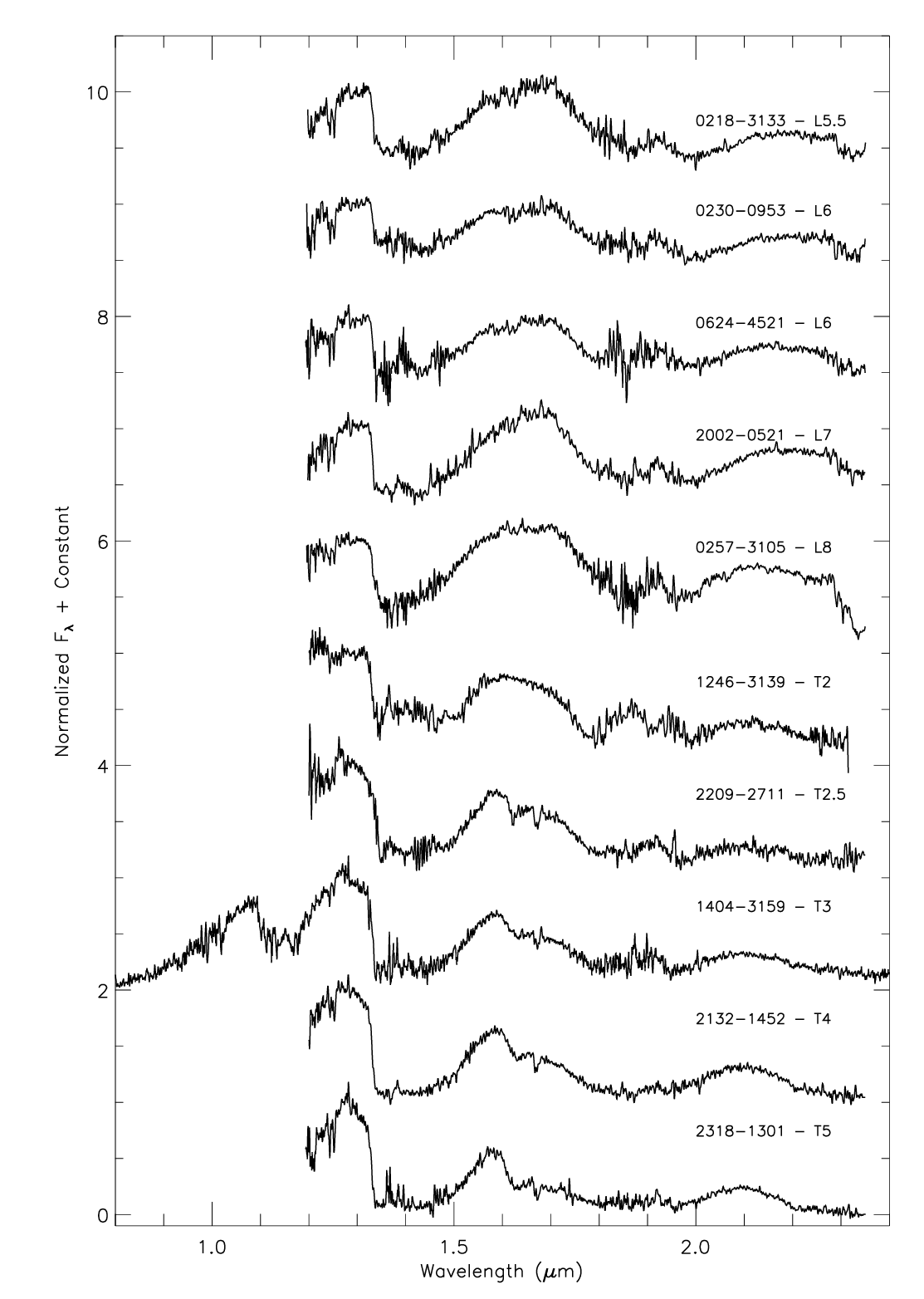}
\caption{Same as Figure \ref{spectra1}, but in the range L5.5-T5. \label{spectra5}}
\end{figure*}

\subsection{Spectral Classification \label{spec_class}}
To determine the spectral types of the objects, all the spectra were fitted to standard template spectra taken from the SpeX-Prism spectral library\footnotemark[2]\footnotetext[2]{http://pono.ucsd.edu/$\sim$adam/browndwarfs/spexprism} using a $\chi^2$ fitting technique, excluding the telluric absorption bands when computing the statistic. The best-fit template was visually inspected to check it was a good fit to the spectrum of the object, and to spot possible peculiarities. If the best-fit template matches the real spectrum, the fit was accepted and the uncertainty on the spectral type of the target was assigned based on a parabolic fit to the $\chi^2$ distribution, rounding it to multiples of 0.5. If the best-fit does not reproduce the spectrum well, we assigned a spectral type to the object based on a ``by-eye'' matching, selecting the template that gives the best match to the J band, and labeling the object as ``peculiar''. The properties of the peculiar objects, their spectral peculiarities, and the assigned spectral types are discussed further in Section \ref{individual}.

A summary of the observations and the results obtained is given in Table \ref{spec}. For each object we list the short name, the instrument used to obtain its spectrum, the night when it was observed, the associated standard and its spectral type, the object's previous optical and NIR classification (if available) and our new NIR spectroscopic classification.

\begin{deluxetable}{l c c c c c c c c}
\tablecaption{Summary of the observations. \label{spec}}
\rotate
\tablewidth{0pt}
\tabletypesize{\scriptsize}
\tablehead{
Object & Instrument & Date of observation & Standard & Standard & Previous & Previous & New & Ref. \\
short name & & (DD-MM-YYYY) & name & type & optical type & NIR type & NIR type & O,I }
\startdata
0014-4844 & OSIRIS   & 2012-09-08 & HIP3471   & A1V & L2.5pec    & \dots & L2.5$\pm$1 & 7,- \\
0032-4405 & OSIRIS   & 2011-08-30 & HD2811    & A3V & L0$\gamma$ & \dots & L4 pec & 1,- \\
0053-3631 & OSIRIS   & 2012-09-08 & HIP6257   & A1V & L3.5       & \dots & L4$\pm$1 & 7,- \\
0058-0651 & OSIRIS   & 2011-09-09 & HIP5164   & A1V & L0         & \dots & L1   & 2,- \\
0109-5100 & OSIRIS   & 2011-09-08 & HIP8241   & A1V & M8.5       & L2    & M7   & 3,3 \\
0128-5545 & OSIRIS   & 2011-09-08 & HIP8241   & A1V & L2         & L1    & L1   & 4,5 \\
0144-4844 & OSIRIS   & 2012-09-08 & HIP10512  & A0V & L5         & \dots & L5   & 16,- \\
0147-4954 & OSIRIS   & 2011-10-06 & HIP8241   & A1V & \dots      & \dots & M9$\pm$1 & \\
0218-3133 & OSIRIS   & 2012-09-10 & HIP12786  & A1V & L3         & \dots & L5.5 & 6,- \\
0219-1939 & OSIRIS   & 2011-09-10 & HD17224   & A0V & L1         & L2.5  & L1$\pm$1 & 3,3 \\
0227-1624 & OSIRIS   & 2012-11-02 & HIP14627  & A0V & L1         & \dots & L0.5$\mp$1 & 4,- \\
0230-0953 & OSIRIS   & 2011-10-06 & HIP10512  & A0V & \dots      & \dots & L6   & \\
0239-1735 & OSIRIS   & 2011-09-10 & HD17224   & A0V & L0         & \dots & M9   & 6,- \\
0257-3105 & OSIRIS   & 2011-10-06 & HR903     & A0V & L8         & \dots & L8$\pm$1 & 7,- \\
0357-4417 & OSIRIS   & 2011-09-10 & HD28813   & A0V & L0$\beta^a$ & \dots & L2 pec$^a$ & 1,- \\
0539-0059 & OSIRIS   & 2011-12-12 & HIP28449  & A0V & L5         & \dots & L4$\pm$1 & 8,- \\
0614-2019 & OSIRIS   & 2011-10-07 & HIP31094  & A0V & \dots      & \dots & L2   & \\
0624-4521 & OSIRIS   & 2012-03-03 & HD47925   & A0V & L5$\pm$1   & \dots & L6   & 4,- \\
0719-5050 & OSIRIS   & 2009-02-14 & HD56980   & A0V & \dots      & \dots & M4   & \\ 
0719-5051 & OSIRIS   & 2011-12-10 & HD60130   & A0V & L0         & \dots & L0$\pm$1 & 4,- \\
0829-1309 & OSIRIS   & 2011-12-12 & HD73687   & A0V & L2         & \dots & L2$\pm$1 & 3,- \\
0832-0128 & OSIRIS   & 2012-03-04 & HR3383    & A1V & L1.5       & \dots & L1$\pm$1 & 2,- \\
0835-0819 & OSIRIS   & 2012-03-04 & HR3383    & A1V & L5         & \dots & L4   & 6,- \\
0909-0658 & OSIRIS   & 2013-01-05 & HIP47249  & A1V & L0         & \dots & L0$\pm$1 & 7,- \\
0928-1603 & OSIRIS   & 2009-02-14 & HIP45800  & A0V & L2         & \dots & L2$\pm$1 & 2,- \\
0953-1014 & OSIRIS   & 2012-03-08 & HD86593   & A0V & L0         & \dots & L2   & 13,- \\
1004-1318 & OSIRIS   & 2012-03-08 & HD86593   & A0V & L0         & \dots & L1$\pm$1 & 14,- \\
1004-3335 & OSIRIS   & 2012-03-09 & HD89213   & A0V & L4         & \dots & L5$\pm$1 & 15,- \\
LHS~5166  & OSIRIS   & 2012-03-09 & HD89213   & A0V & \dots      & \dots & M4   & \\
1045-0149 & OSIRIS   & 2013-01-05 & HIP54849  & A0V & L1         & \dots & L2   & 15,- \\
1059-2113 & OSIRIS   & 2013-03-17 & HIP55830  & A0V & L1         & \dots & L2   & 6,- \\
1154-3400 & OSIRIS   & 2013-03-07 & HIP61211  & A0V & L0         & \dots & L3   & 7,- \\
1246-3139 & OSIRIS   & 2011-02-25 & HIP60819  & A0V & \dots      & \dots & T2   & \\
1331-0116 & XSHOOTER & 2011-06-07 & HIP68713  & A0V & L6         & L8$\pm$2.5 & L1 pec & 9,10 \\
1404-3159 & XSHOOTER & 2011-06-05 & HIP065688 & B8V & T0$^a$     & T2.5$^a$ & T3$^a$ & 11,12 \\
1438-1309 & OSIRIS   & 2012-06-04 & HD132072  & A0V & L3$\pm$1   & \dots & L3$\pm$1 & 2,- \\
1753-6559 & SOFI     & 2011-04-21 & HD168741  & A0V & L4$\pm$2   & \dots & L4$\pm$1 & 4,- \\
1928-4356 & OSIRIS   & 2009-06-08 & HIP95464  & A0V & L4         & \dots & L4 pec & 4,- \\
1936-5502 & SOFI     & 2011-04-21 & HD168741  & A0V & L5$\pm$1   & \dots & L4   & 4,- \\
2002-0521 & OSIRIS   & 2012-06-04 & HIP98953  & A0V & L6         & \dots & L7   & 13,- \\
2011-6201 & XSHOOTER & 2011-06-05 & HD192510  & A0V & \dots      & \dots & d/sdM8 & \\
2023-5946 & XSHOOTER & 2011-06-05 & HD192510  & A0V & \dots      & \dots & M8   & \\
2045-6332 & OSIRIS   & 2011-09-10 & HD197165  & A3V & \dots      & \dots & L1$\pm$1 & \\
2101-2944 & OSIRIS   & 2011-04-29 & HIP103315 & A0V & \dots      & \dots & L1   & \\
2132-1452 & OSIRIS   & 2011-08-30 & HD206703  & A3V & \dots      & \dots & T4   & \\
2148-6323 & XSHOOTER & 2011-06-06 & HIP097611 & B5V & \dots      & \dots & L1   & \\
2158-1550 & OSIRIS   & 2012-06-04 & HD211278  & A0V & L4$\pm$1   & \dots & L5 & 7,- \\
2209-2711 & OSIRIS   & 2011-06-11 & HD211278  & A0V & \dots      & \dots & T2.5 & \\
2213-2136 & OSIRIS   & 2011-08-30 & HR8542    & A0V & L0$\gamma$ & \dots & L2 pec & 1,- \\
2310-1759 & OSIRIS   & 2011-09-08 & HD219179  & A3V & L0$\pm$1   & L1    & L1$\pm$1 & 13,3 \\
2318-1301 & OSIRIS   & 2012-09-16 & HIP117734 & A1V & \dots      & \dots & T5   & \\
2346-5928 & OSIRIS   & 2011-09-08 & HD224377  & A0V & \dots      & \dots & M7 pec & \\
\enddata
\tablecomments{For each object we present the instrument used to obtain its spectrum, the date of observation, the telluric standard used and its spectral type, the previous optical and NIR classification of the target, our new NIR spectral classification, and the references to the previous types (optical and NIR). If not specified, the uncertainty on the new NIR type is 0.5. ($a$) Known unresolved binary, the reported type is the unresolved classification. The spectral types of the components are determined and discussed further in Section \ref{unres_bins}.} 
\tablerefs{(1) \citet{2009AJ....137.3345C}; (2) \citet{2000AJ....120..447K}; (3) \citet{2005A&A...440.1061L}; (4) \citet{2008AJ....136.1290R}; (5) \citet{2007MNRAS.374..445K}; (6) \citet{2003AJ....126.2421C}; (7) \citet{2008ApJ...689.1295K}; (8) \citet{2000AJ....119..928F}; (9) \citet{2002AJ....123.3409H}; (10) \citet{2004AJ....127.3553K}; (11) \citet{2008ApJ...685.1183L}; (12) \citet{2007AJ....134.1162L}; (13) \citet{2007AJ....133..439C}; (14) \citet{2010A&A...517A..53M}; (15) \citet{2002ApJ...575..484G}; (16) \citet{2003AJ....125..343L}.}
\end{deluxetable}

\begin{figure}
\includegraphics[width=7.6cm,height=16cm]{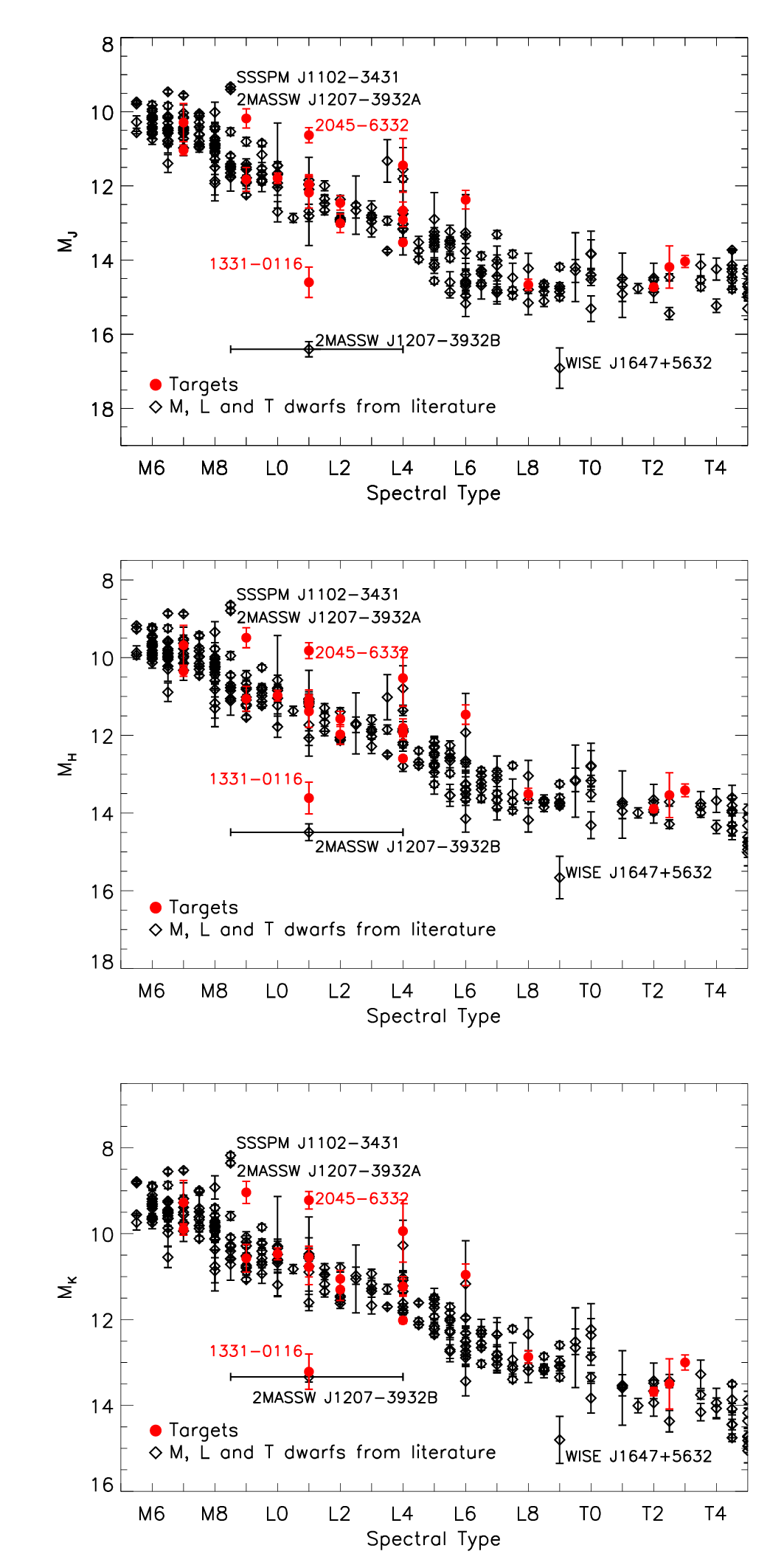}
\caption{Absolute 2MASS JHK magnitudes as a function of spectral type. The objects presented in this paper are plotted as red circles. Other objects are plotted as black diamonds. Magnitudes and parallaxes for the ultracool dwarfs are taken from \citet{2012ApJS..201...19D}. For literature objects, we plot the NIR spectral type when available, otherwise we use the optical spectral type. The outliers are marked, and more details about them can be found in the text (Section \ref{spec_class}).}
\label{hr}
\end{figure}

In Figure \ref{hr} we plot the absolute 2MASS JHK$_s$ magnitudes as a function of spectral type. The targets presented here are plotted as red dots, while diamonds represent objects taken from the literature \citep[see][Table 9, for a complete census of ultracool dwarfs with measured parallaxes]{2012ApJS..201...19D}. For the literature sample, we use the NIR spectral type when available, otherwise we plot the optical spectral type. Our sample represents a significant increase in the number of objects with measured parallaxes and NIR spectral types at early types (L0-L4). Most of the previous parallax programs have indeed focused on the cooler, later-type targets.

We notice that the scatter in absolute magnitudes across the sequence is $\sim$1 mag on average but goes up to $\sim$2 mag between late-M and mid-L. The number of objects per spectral type, especially at early types, is high enough to allow us to identify the outliers (which are marked in Figure \ref{hr}), and therefore the remaining scatter is most likely intrinsic to the sequence. The cause of such a spread can probably be found in the known mass-age degeneracy, typical of brown dwarfs. Objects of the same spectral type can have very different mass and age, and this would result in peculiar colours, but also in differences in the absolute magnitude of the dwarfs \citep[e.g.][and references therein]{2005ARA&A..43..195K}.

There are six outliers to the sequence, and they are marked in Figure \ref{hr}. Two are the components of the brown dwarf + planet system 2MASSW~J1207334$-$393254. This system is part of the TW Hydrae Association, with an age of 8$^{+4}_{-3}$ Myr \citep[][and references therein]{2004A&A...425L..29C}. The primary is a M8.5 dwarf \citep{2002ApJ...575..484G}, while for the planetary companion \citet{2010A&A...517A..76P} derived a spectral type in the range M8.5$-$L4. The primary is $\sim$1 mag overluminous compared to objects of similar spectral type, as expected for an object that has not contracted to its final radius \citep[e.g.][]{1997ApJ...491..856B}. On the other hand, the companion is more than 1 mag underluminous compared to objects of similar spectral type (i.e. in the M8.5$-$L4 range) and $\sim$2.5 mag underluminous when compared to standard models of giant planet evolution \citep{2011ApJ...735L..39B,2011ApJ...732..107S}. \citet{2012ApJ...752...56F} have found similar results for other young, very red L dwarfs, and have speculated that the underluminosity can be due to a combination of two factors. One is the possibility that the low-gravity spectral classification have a different temperature relation compared to the standard classification scheme \citep{1999ApJ...519..802K,2006ApJ...637.1067B}. The other factor is the possibility that young L dwarfs have dustier photospheres, that make them appear fainter and redder in the NIR compared to other field L dwarfs. SSSPM~J1102$-$3431 is also $\sim$1 mag overluminous respect to the other M8.5 plotted. This object is another young M dwarfs, known to be part of the TW Hydrae Association \citep{2005A&A...430L..49S,2008A&A...489..825T}. Another outlier is the peculiar red L9 dwarf WISEPA~J164715.59+563208.2 \citep{2011ApJS..197...19K}. This object pertains to the class of peculiar red, non-low-gravity L dwarfs, whose nature is not yet fully understood \citep[e.g.][]{2010ApJS..190..100K}. Finally, the last two outliers are our targets 1331-0116 and 2045-6332. The first one, 1331-0116, is a peculiar blue L1 dwarf, while 2045-6332 is overluminous and very red (J-K$_s$ = 1.41). We discuss further their properties in Section \ref{sdss1331} and \ref{2045-6332}.

\begin{figure}
\includegraphics[width=7.6cm,height=16cm]{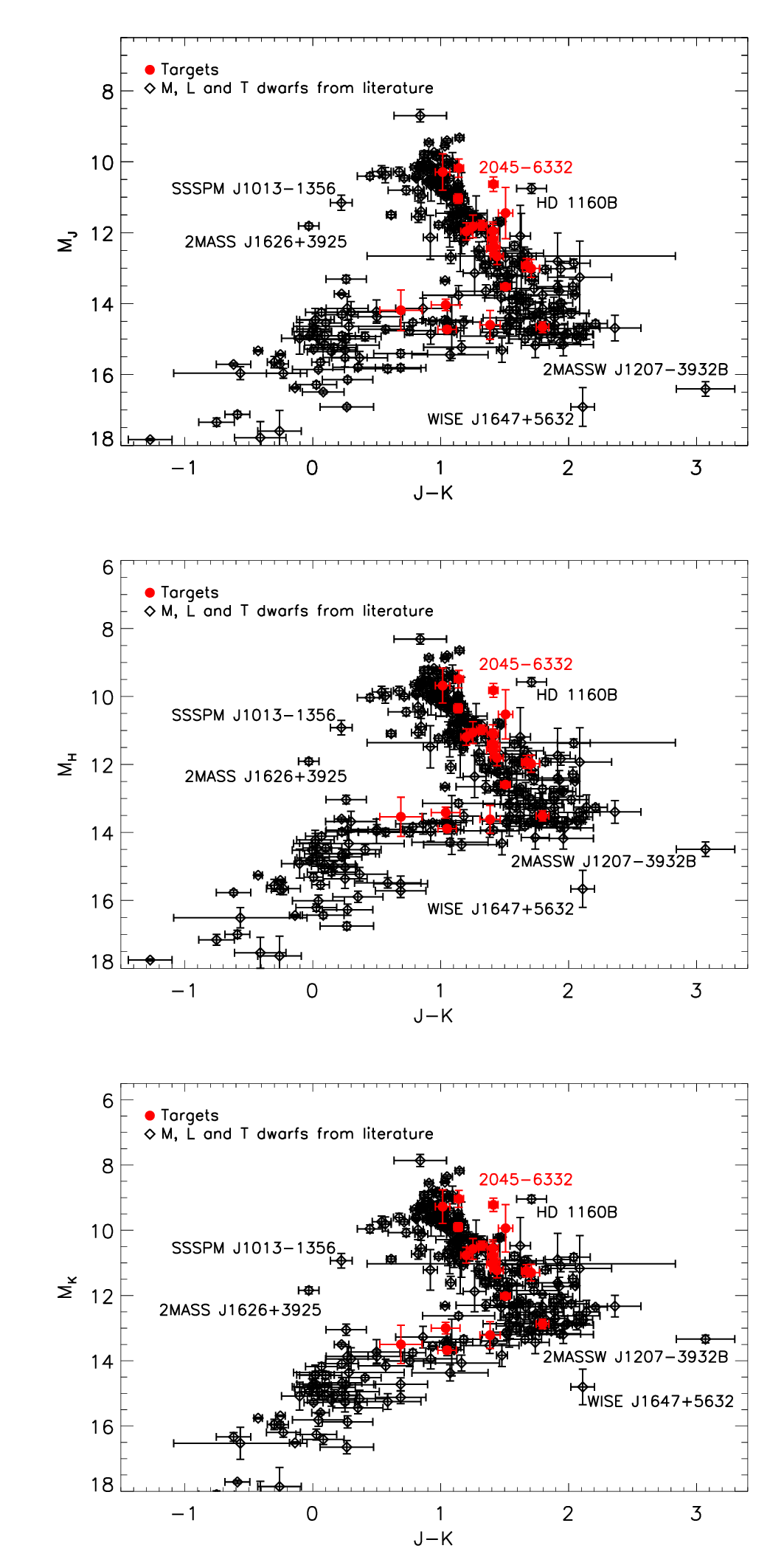}
\caption{Absolute 2MASS JHK magnitudes as a function of J-K. Colours and symbols follow the same convention as in Figure \ref{hr}. Magnitudes and parallaxes for the ultracool dwarfs are taken from \citet{2012ApJS..201...19D}. The outliers are marked, and more details about them can be found in the text (Section \ref{spec_class}).}
\label{colour-mag}
\end{figure}

To better understand the properties of the objects in our sample, in Figure \ref{colour-mag} we plot the absolute 2MASS JHK$_s$ magnitudes as a function of J-K$_s$. Colours and symbols follow the same convention as in Figure \ref{hr}.

The objects follow the expected trend, moving towards redder colours as they become fainter because of the thickening of the clouds deck in their atmospheres. Brown dwarfs then rapidly turn towards blue colours at the L-T transition, because of the dust settling and the onset of the CH$_4$ and CIA absorption. The colour turnaround is sharper in the top panel of Figure \ref{colour-mag} ($M_J$ vs. J-K) but also shows a larger scatter in absolute magnitude (almost 2 mag) compared to the bottom panel ($M_K$ vs. J-K) where the transition is shallower but the scatter is only $\sim$1 mag.

The outliers are labelled. Two are the already mentioned 2MASSW~J1207334$-$393254B and WISEPA~J164715.59+563208.2. Another outlier is the very red HD~1160B, a L dwarf companion to the young star HD~1160A \citep{2012ApJ...750...53N}. As stated above, the unusual red colours of these objects can be explained assuming an enhanced dust content in their atmospheres \citep[e.g.][]{2012ApJ...752...56F}. Another two outliers are the unusually blue SSSPM~J1013$-$1356 and 2MASS~J16262034+3925190, both known to be subdwarfs \citep[sdM9.5 and sdL4 respectively,][]{2009AJ....137....1F}. Their blue colour are most probably due to the reduced cloud opacity (which affects especially the J-band) due to their low metallicity \citep[e.g.][]{2012ApJ...752...56F}. Finally, the last outlier is our target 2045-6332, which is the most luminous L1. Its very red J-K$_s$ could be an indication of low surface gravity, therefore of youth. We discuss further the nature of this object in Section \ref{2045-6332}.

\begin{figure*}
\includegraphics[width=16cm]{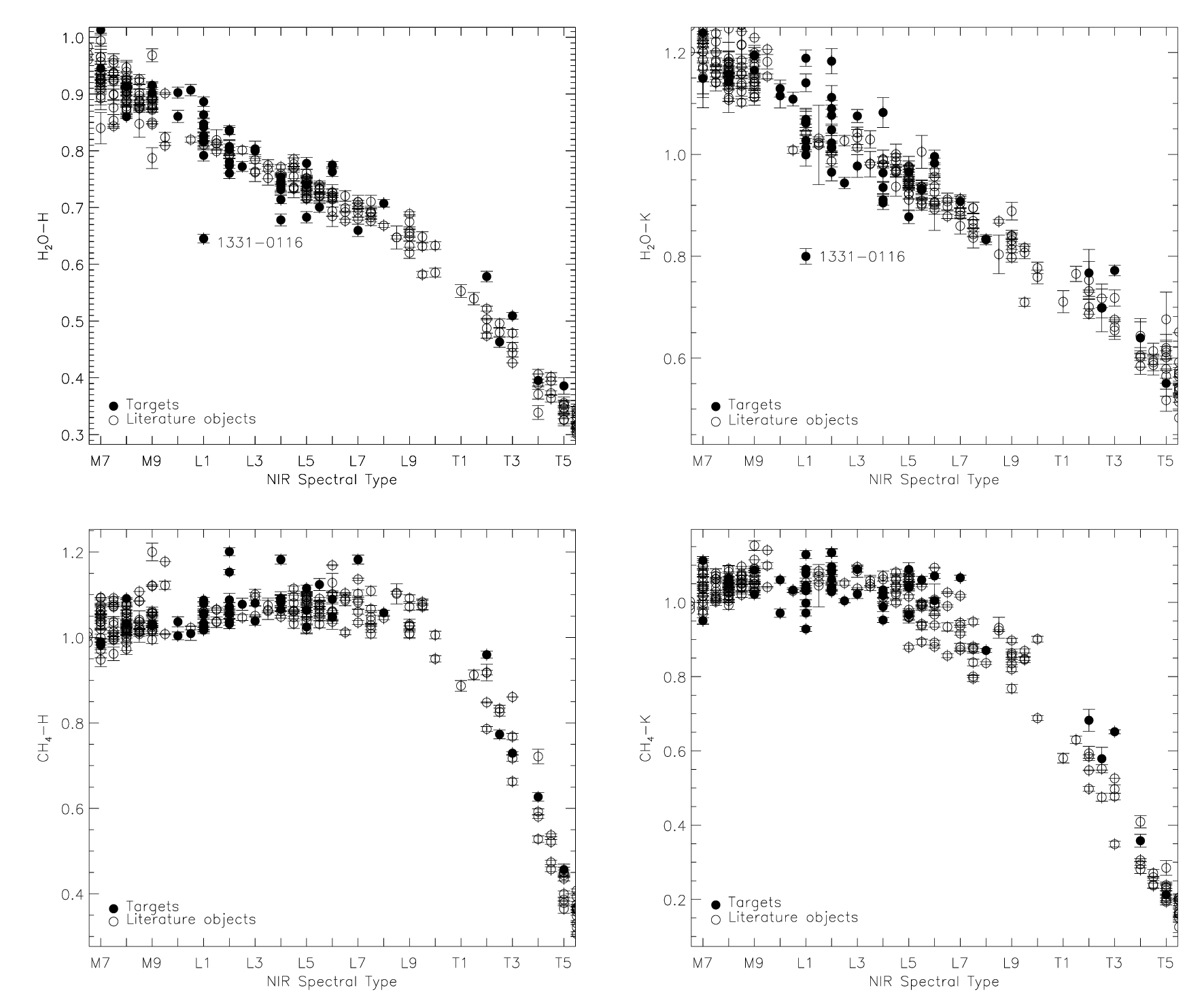}
\caption{The spectral indices derived for our targets as a function of their spectral type. The indices presented here are defined in \citet{2006ApJ...637.1067B} and \citet{2010ApJ...710.1142B}.}
\label{indices_plot}
\end{figure*}

In Table \ref{indices} we present the spectral indices calculated for our targets. The indices are those defined in \citet{2006ApJ...637.1067B} and \citet{2010ApJ...710.1142B}. The indices are plotted also in Figure \ref{indices_plot}, where we show H2O-H and H2O-K as a function of spectral type (top two panels) and CH4-H and CH4-K as a function of spectral type. Our targets are plotted as filled circles, while literature objects are overplotted as open circles for comparison. The literature objects are taken from the SpeX-Prism library. We can see that the H2O-H and H2O-K correlate very well with the fit-based spectral type, with only one outstanding outlier to the sequence, which is once again 1331-0116. Its indices have values that are typical of much later type objects (L8-T0) because of the unusually strong H$_2$O absorption bands showed by its spectrum. The CH4-H and CH4-K indices correlates with the spectral type only for the late-L and T dwarfs (i.e. from L8 onward). We also note that the scatter is larger compared to the H$_2$O indices. The only outlier to the  sequence is 1404-3159, whose unusual position in the bottom right plot of Figure \ref{indices_plot} is due to its binarity. In the $K$-band, the early type component of the binary dominates, and its methane absorption is less prominent than in the late type component, and also weaker than it would be in a single object of type T3. The discrepancy is not present in the CH4-H plot as in the $H$-band the contribution to the total flux coming from the two components is almost equal (similarly for the H2O-H and H2O-K plots).

\begin{deluxetable}{l c c c c c c c c}
\tablecaption{Spectral indices for the objects in the sample. \label{indices}}
\rotate
\tablewidth{0pt}
\tabletypesize{\scriptsize}
\tablehead{
Object & H2O-J & H2O-H & H2O-K & CH4-J & CH4-H & CH4-K & K/J & H-dip \\
short name & & & & & & & & }
\startdata
0014-4844  & \dots & 0.772 $\pm$ 0.009 & 0.944 $\pm$ 0.011 & 0.862 $\pm$ 0.016 & 1.078 $\pm$ 0.014 & 1.004 $\pm$ 0.005 & 0.461 $\pm$ 0.005 & 0.491 $\pm$ 0.006 \\
0032-4405  & \dots & 0.678 $\pm$ 0.010 & 1.082 $\pm$ 0.029 & 0.998 $\pm$ 0.015 & 1.182 $\pm$ 0.011 & 1.026 $\pm$ 0.014 & 0.486 $\pm$ 0.006 & 0.502 $\pm$ 0.005 \\
0053-3631  & \dots & 0.747 $\pm$ 0.009 & 0.935 $\pm$ 0.012 & 0.860 $\pm$ 0.015 & 1.091 $\pm$ 0.016 & 1.033 $\pm$ 0.007 & 0.526 $\pm$ 0.006 & 0.495 $\pm$ 0.007 \\
0058-0651  & \dots & 0.825 $\pm$ 0.007 & 1.062 $\pm$ 0.023 & 0.876 $\pm$ 0.011 & 1.053 $\pm$ 0.007 & 1.129 $\pm$ 0.011 & 0.434 $\pm$ 0.005 & 0.486 $\pm$ 0.004 \\
0109-5100  & \dots & 1.013 $\pm$ 0.006 & 1.239 $\pm$ 0.020 & 0.872 $\pm$ 0.010 & 0.982 $\pm$ 0.006 & 1.113 $\pm$ 0.005 & 0.311 $\pm$ 0.002 & 0.481 $\pm$ 0.004 \\
0128-5545  & \dots & 0.842 $\pm$ 0.008 & 1.063 $\pm$ 0.026 & 0.875 $\pm$ 0.015 & 1.017 $\pm$ 0.008 & 0.971 $\pm$ 0.008 & 0.473 $\pm$ 0.005 & 0.486 $\pm$ 0.005 \\
0144-0716  & \dots & 0.778 $\pm$ 0.011 & 0.966 $\pm$ 0.010 & 0.870 $\pm$ 0.014 & 1.025 $\pm$ 0.016 & 0.967 $\pm$ 0.006 & 0.623 $\pm$ 0.006 & 0.477 $\pm$ 0.007 \\
0147-4954  & \dots & 0.915 $\pm$ 0.004 & 1.165 $\pm$ 0.017 & 0.963 $\pm$ 0.013 & 1.028 $\pm$ 0.005 & 1.087 $\pm$ 0.005 & 0.424 $\pm$ 0.002 & 0.497 $\pm$ 0.003 \\
0218-3133  & \dots & 0.701 $\pm$ 0.009 & 0.930 $\pm$ 0.018 & 0.887 $\pm$ 0.017 & 1.124 $\pm$ 0.014 & 1.061 $\pm$ 0.010 & 0.565 $\pm$ 0.007 & 0.487 $\pm$ 0.006 \\
0219-1939  & \dots & 0.817 $\pm$ 0.008 & 1.027 $\pm$ 0.018 & 0.846 $\pm$ 0.012 & 1.022 $\pm$ 0.007 & 1.036 $\pm$ 0.008 & 0.370 $\pm$ 0.004 & 0.484 $\pm$ 0.004 \\
0227-1624  & \dots & 0.907 $\pm$ 0.010 & 1.108 $\pm$ 0.013 & 0.968 $\pm$ 0.012 & 1.009 $\pm$ 0.016 & 1.033 $\pm$ 0.006 & 0.458 $\pm$ 0.004 & 0.491 $\pm$ 0.007 \\
0230-0953  & \dots & 0.775 $\pm$ 0.006 & 0.982 $\pm$ 0.010 & 0.924 $\pm$ 0.014 & 1.049 $\pm$ 0.009 & 1.071 $\pm$ 0.006 & 0.648 $\pm$ 0.006 & 0.486 $\pm$ 0.005 \\
0239-1735  & \dots & 0.902 $\pm$ 0.006 & 1.194 $\pm$ 0.020 & 0.924 $\pm$ 0.011 & 1.032 $\pm$ 0.006 & 1.021 $\pm$ 0.010 & 0.421 $\pm$ 0.004 & 0.495 $\pm$ 0.003 \\
0257-3105  & \dots & 0.707 $\pm$ 0.006 & 0.833 $\pm$ 0.010 & 0.835 $\pm$ 0.014 & 1.058 $\pm$ 0.006 & 0.870 $\pm$ 0.005 & 0.731 $\pm$ 0.006 & 0.515 $\pm$ 0.003 \\
0357-4417  & \dots & 0.780 $\pm$ 0.006 & 1.090 $\pm$ 0.016 & 0.884 $\pm$ 0.013 & 1.153 $\pm$ 0.006 & 1.134 $\pm$ 0.010 & 0.616 $\pm$ 0.005 & 0.498 $\pm$ 0.003 \\
0539-0059  & \dots & 0.714 $\pm$ 0.007 & 0.905 $\pm$ 0.013 & 0.867 $\pm$ 0.018 & 1.081 $\pm$ 0.010 & 0.952 $\pm$ 0.007 & 0.485 $\pm$ 0.005 & 0.490 $\pm$ 0.005 \\
0614-2019  & \dots & 0.802 $\pm$ 0.010 & 1.022 $\pm$ 0.019 & 0.941 $\pm$ 0.018 & 1.070 $\pm$ 0.011 & 1.096 $\pm$ 0.010 & 0.468 $\pm$ 0.006 & 0.487 $\pm$ 0.006 \\
0624-4521  & \dots & 0.763 $\pm$ 0.008 & 0.996 $\pm$ 0.012 & 0.830 $\pm$ 0.021 & 1.089 $\pm$ 0.007 & 1.005 $\pm$ 0.007 & 0.721 $\pm$ 0.008 & 0.491 $\pm$ 0.003 \\
0719-5051A & \dots & 1.057 $\pm$ 0.003 & 1.282 $\pm$ 0.014 & 0.944 $\pm$ 0.003 & 0.968 $\pm$ 0.002 & 0.910 $\pm$ 0.002 & 0.282 $\pm$ 0.001 & 0.503 $\pm$ 0.001 \\
0719-5051B & \dots & 0.902 $\pm$ 0.009 & 1.115 $\pm$ 0.024 & 0.881 $\pm$ 0.015 & 1.004 $\pm$ 0.009 & 0.971 $\pm$ 0.011 & 0.381 $\pm$ 0.005 & 0.482 $\pm$ 0.005 \\
0829-1309  & \dots & 0.837 $\pm$ 0.007 & 1.013 $\pm$ 0.013 & 0.918 $\pm$ 0.014 & 1.032 $\pm$ 0.011 & 1.028 $\pm$ 0.006 & 0.465 $\pm$ 0.004 & 0.482 $\pm$ 0.006 \\
0832-0128  & \dots & 0.847 $\pm$ 0.009 & 1.061 $\pm$ 0.014 & 0.854 $\pm$ 0.015 & 1.057 $\pm$ 0.007 & 1.045 $\pm$ 0.012 & 0.468 $\pm$ 0.005 & 0.485 $\pm$ 0.004 \\
0835-0819  & \dots & 0.753 $\pm$ 0.008 & 0.964 $\pm$ 0.018 & 0.843 $\pm$ 0.016 & 1.092 $\pm$ 0.008 & 0.988 $\pm$ 0.013 & 0.476 $\pm$ 0.006 & 0.498 $\pm$ 0.005 \\
0909-0658  & \dots & 0.860 $\pm$ 0.011 & 1.129 $\pm$ 0.017 & 0.929 $\pm$ 0.016 & 1.037 $\pm$ 0.012 & 1.060 $\pm$ 0.008 & 0.413 $\pm$ 0.004 & 0.494 $\pm$ 0.008 \\
0928-1603  & \dots & 0.774 $\pm$ 0.013 & 1.077 $\pm$ 0.020 & 0.935 $\pm$ 0.024 & 1.055 $\pm$ 0.010 & 1.062 $\pm$ 0.008 & 0.527 $\pm$ 0.009 & 0.484 $\pm$ 0.005 \\
0953-1014  & \dots & 0.807 $\pm$ 0.012 & 1.112 $\pm$ 0.023 & 0.911 $\pm$ 0.014 & 1.088 $\pm$ 0.015 & 1.134 $\pm$ 0.010 & 0.503 $\pm$ 0.005 & 0.502 $\pm$ 0.006 \\
1004-1318  & \dots & 0.792 $\pm$ 0.010 & 0.999 $\pm$ 0.022 & 0.820 $\pm$ 0.022 & 1.026 $\pm$ 0.009 & 0.998 $\pm$ 0.009 & 0.474 $\pm$ 0.005 & 0.483 $\pm$ 0.005 \\
1004-3335  & \dots & 0.740 $\pm$ 0.012 & 0.965 $\pm$ 0.031 & 0.851 $\pm$ 0.017 & 1.064 $\pm$ 0.015 & 1.088 $\pm$ 0.019 & 0.561 $\pm$ 0.009 & 0.478 $\pm$ 0.007 \\
LHS5166    & \dots & 1.075 $\pm$ 0.004 & 1.334 $\pm$ 0.019 & 0.925 $\pm$ 0.005 & 0.954 $\pm$ 0.003 & 0.919 $\pm$ 0.003 & 0.299 $\pm$ 0.001 & 0.500 $\pm$ 0.002 \\
1045-0149  & \dots & 0.835 $\pm$ 0.006 & 1.048 $\pm$ 0.011 & 0.930 $\pm$ 0.012 & 1.072 $\pm$ 0.010 & 1.081 $\pm$ 0.006 & 0.529 $\pm$ 0.005 & 0.485 $\pm$ 0.005 \\
1059-2113  & \dots & 0.760 $\pm$ 0.009 & 0.965 $\pm$ 0.017 & 0.872 $\pm$ 0.016 & 1.036 $\pm$ 0.007 & 1.040 $\pm$ 0.010 & 0.487 $\pm$ 0.006 & 0.489 $\pm$ 0.004 \\
1154-3400  & \dots & 0.800 $\pm$ 0.006 & 1.075 $\pm$ 0.013 & 0.920 $\pm$ 0.023 & 1.080 $\pm$ 0.007 & 1.089 $\pm$ 0.007 & 0.473 $\pm$ 0.004 & 0.498 $\pm$ 0.004 \\
1246-3139  & \dots & 0.578 $\pm$ 0.009 & 0.767 $\pm$ 0.046 & 0.747 $\pm$ 0.021 & 0.960 $\pm$ 0.008 & 0.682 $\pm$ 0.030 & 0.376 $\pm$ 0.005 & 0.529 $\pm$ 0.003 \\
1331-0116  & 0.721 $\pm$ 0.010 & 0.645 $\pm$ 0.007 & 0.800 $\pm$ 0.015 & 0.713 $\pm$ 0.010 & 1.079 $\pm$ 0.006 & 0.928 $\pm$ 0.006 & 0.375 $\pm$ 0.004 & 0.512 $\pm$ 0.003 \\
1404-3159  & 0.444 $\pm$ 0.006 & 0.509 $\pm$ 0.006 & 0.772 $\pm$ 0.010 & 0.554 $\pm$ 0.006 & 0.729 $\pm$ 0.003 & 0.651 $\pm$ 0.005 & 0.311 $\pm$ 0.002 & 0.453 $\pm$ 0.002 \\
1438-1309  & \dots & 0.803 $\pm$ 0.014 & 0.977 $\pm$ 0.022 & 0.868 $\pm$ 0.017 & 1.038 $\pm$ 0.010 & 1.022 $\pm$ 0.012 & 0.538 $\pm$ 0.008 & 0.480 $\pm$ 0.006 \\
1753-6559  & 0.843 $\pm$ 0.015 & 0.741 $\pm$ 0.014 & \dots             & 0.888 $\pm$ 0.019 & \dots             & \dots             & \dots             & \dots \\
1928-4356  & \dots & 0.735 $\pm$ 0.007 & 0.911 $\pm$ 0.011 & 0.858 $\pm$ 0.020 & 1.066 $\pm$ 0.005 & 1.018 $\pm$ 0.006 & 0.400 $\pm$ 0.004 & 0.484 $\pm$ 0.003 \\
1936-5502  & 0.757 $\pm$ 0.015 & 0.732 $\pm$ 0.009 & \dots             & 0.845 $\pm$ 0.024 & \dots             & \dots             & \dots             & \dots \\
2002-0521  & \dots & 0.660 $\pm$ 0.011 & 0.908 $\pm$ 0.013 & 0.815 $\pm$ 0.018 & 1.182 $\pm$ 0.011 & 1.066 $\pm$ 0.007 & 0.720 $\pm$ 0.008 & 0.504 $\pm$ 0.005 \\
2011-6201  & 0.993 $\pm$ 0.004 & 0.860 $\pm$ 0.003 & 1.142 $\pm$ 0.008 & 0.840 $\pm$ 0.005 & 1.091 $\pm$ 0.003 & 1.065 $\pm$ 0.004 & 0.281 $\pm$ 0.001 & 0.504 $\pm$ 0.001 \\
2023-5946  & 1.016 $\pm$ 0.002 & 0.913 $\pm$ 0.002 & 1.158 $\pm$ 0.004 & 0.913 $\pm$ 0.002 & 1.037 $\pm$ 0.001 & 1.049 $\pm$ 0.002 & 0.360 $\pm$ 0.001 & 0.498 $\pm$ 0.001 \\
2045-6332  & \dots & 0.863 $\pm$ 0.006 & 1.189 $\pm$ 0.016 & 0.895 $\pm$ 0.011 & 1.087 $\pm$ 0.006 & 1.089 $\pm$ 0.007 & 0.394 $\pm$ 0.003 & 0.499 $\pm$ 0.003 \\
2101-2944  & \dots & 0.817 $\pm$ 0.010 & 1.069 $\pm$ 0.022 & 0.957 $\pm$ 0.017 & 1.033 $\pm$ 0.010 & 1.077 $\pm$ 0.012 & 0.360 $\pm$ 0.005 & 0.501 $\pm$ 0.005 \\
2132-1452  & \dots & 0.395 $\pm$ 0.011 & 0.640 $\pm$ 0.031 & 0.550 $\pm$ 0.019 & 0.627 $\pm$ 0.010 & 0.358 $\pm$ 0.017 & 0.312 $\pm$ 0.005 & 0.441 $\pm$ 0.007 \\
2148-6323  & 0.895 $\pm$ 0.007 & 0.827 $\pm$ 0.004 & 1.013 $\pm$ 0.008 & 0.890 $\pm$ 0.006 & 1.018 $\pm$ 0.005 & 1.034 $\pm$ 0.005 & 0.446 $\pm$ 0.002 & 0.479 $\pm$ 0.003 \\
2158-1550  & \dots & 0.683 $\pm$ 0.010 & 0.877 $\pm$ 0.013 & 0.827 $\pm$ 0.016 & 1.115 $\pm$ 0.007 & 1.041 $\pm$ 0.007 & 0.587 $\pm$ 0.006 & 0.486 $\pm$ 0.005 \\
2209-2711  & \dots & 0.463 $\pm$ 0.009 & 0.699 $\pm$ 0.047 & 0.704 $\pm$ 0.019 & 0.773 $\pm$ 0.010 & 0.579 $\pm$ 0.031 & 0.261 $\pm$ 0.007 & 0.448 $\pm$ 0.009 \\
2213-2136  & \dots & 0.805 $\pm$ 0.012 & 1.183 $\pm$ 0.025 & 0.973 $\pm$ 0.015 & 1.201 $\pm$ 0.009 & 1.086 $\pm$ 0.012 & 0.637 $\pm$ 0.007 & 0.519 $\pm$ 0.004 \\
2310-1759  & \dots & 0.886 $\pm$ 0.008 & 1.140 $\pm$ 0.017 & 0.899 $\pm$ 0.013 & 1.026 $\pm$ 0.007 & 1.031 $\pm$ 0.008 & 0.423 $\pm$ 0.004 & 0.490 $\pm$ 0.004 \\
2318-1301  & \dots & 0.386 $\pm$ 0.015 & 0.551 $\pm$ 0.025 & 0.484 $\pm$ 0.019 & 0.457 $\pm$ 0.013 & 0.212 $\pm$ 0.017 & 0.243 $\pm$ 0.004 & 0.348 $\pm$ 0.009 \\
2346-5928  & \dots & 0.945 $\pm$ 0.012 & 1.149 $\pm$ 0.030 & 0.843 $\pm$ 0.011 & 0.990 $\pm$ 0.008 & 0.951 $\pm$ 0.008 & 0.265 $\pm$ 0.002 & 0.493 $\pm$ 0.004 \\
\enddata
\tablecomments{The indices presented here are defined in \citet{2006ApJ...637.1067B} and \citet{2010ApJ...710.1142B}.}
\end{deluxetable}

\subsection{Unresolved Binaries \label{unres_bins}}
We also performed a search for unresolved binaries within our sample, using the spectral indices and the criteria defined by \citet{2010ApJ...710.1142B}. Two of the objects in the sample are known unresolved binaries: 1404-3159 \citep{2008ApJ...685.1183L} is indeed identified as a strong candidate by the selection criteria; 0357-4417 \citep{2003AJ....126.1526B} is not selected as a candidate, a result which is not surprising, as this is an early-L pair, and the technique used is sensitive mostly to L-T transition systems. None of the other objects in our sample match the criteria defined. 

\begin{figure}
\includegraphics[width=7.6cm]{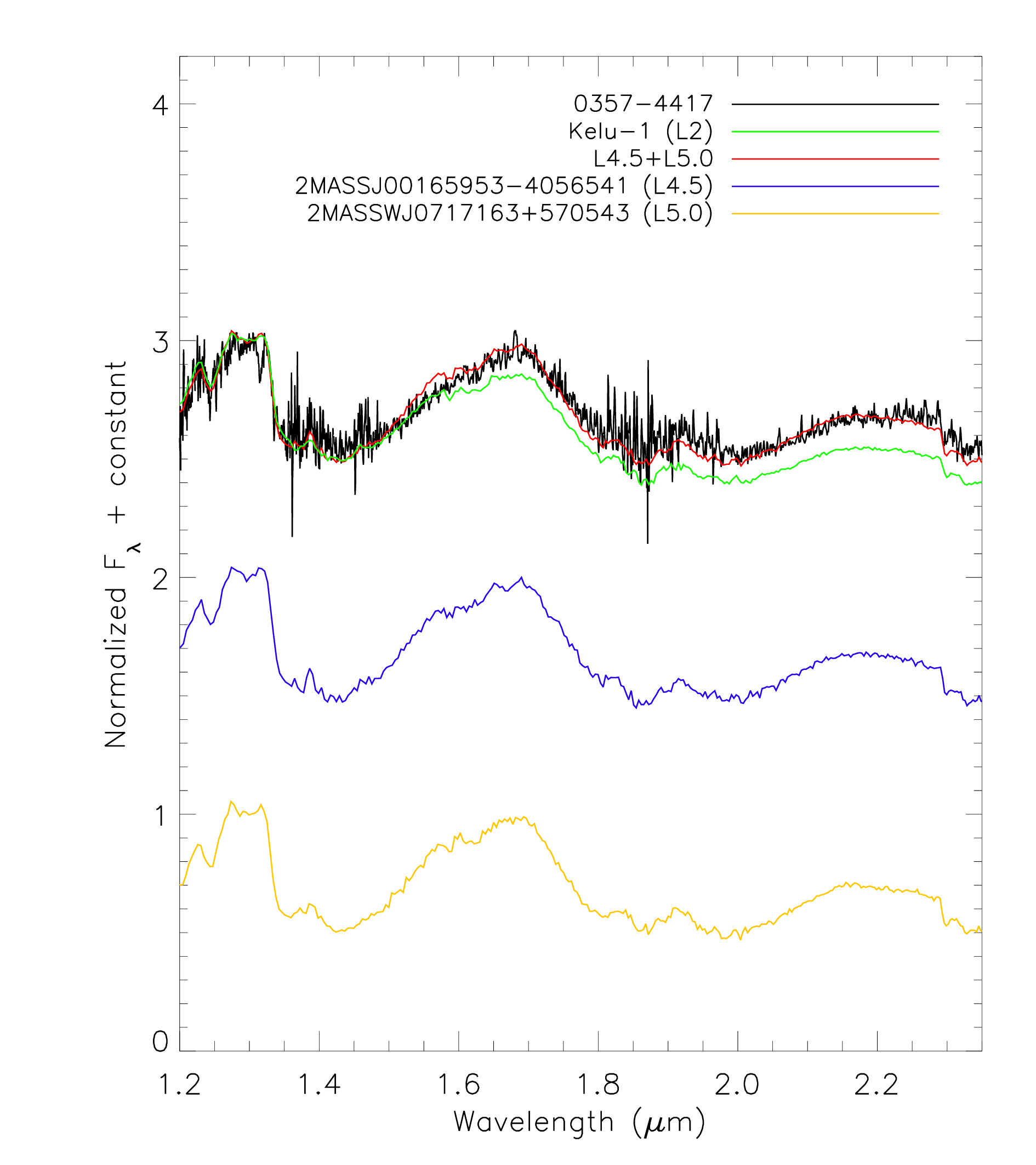}
\includegraphics[width=7.6cm]{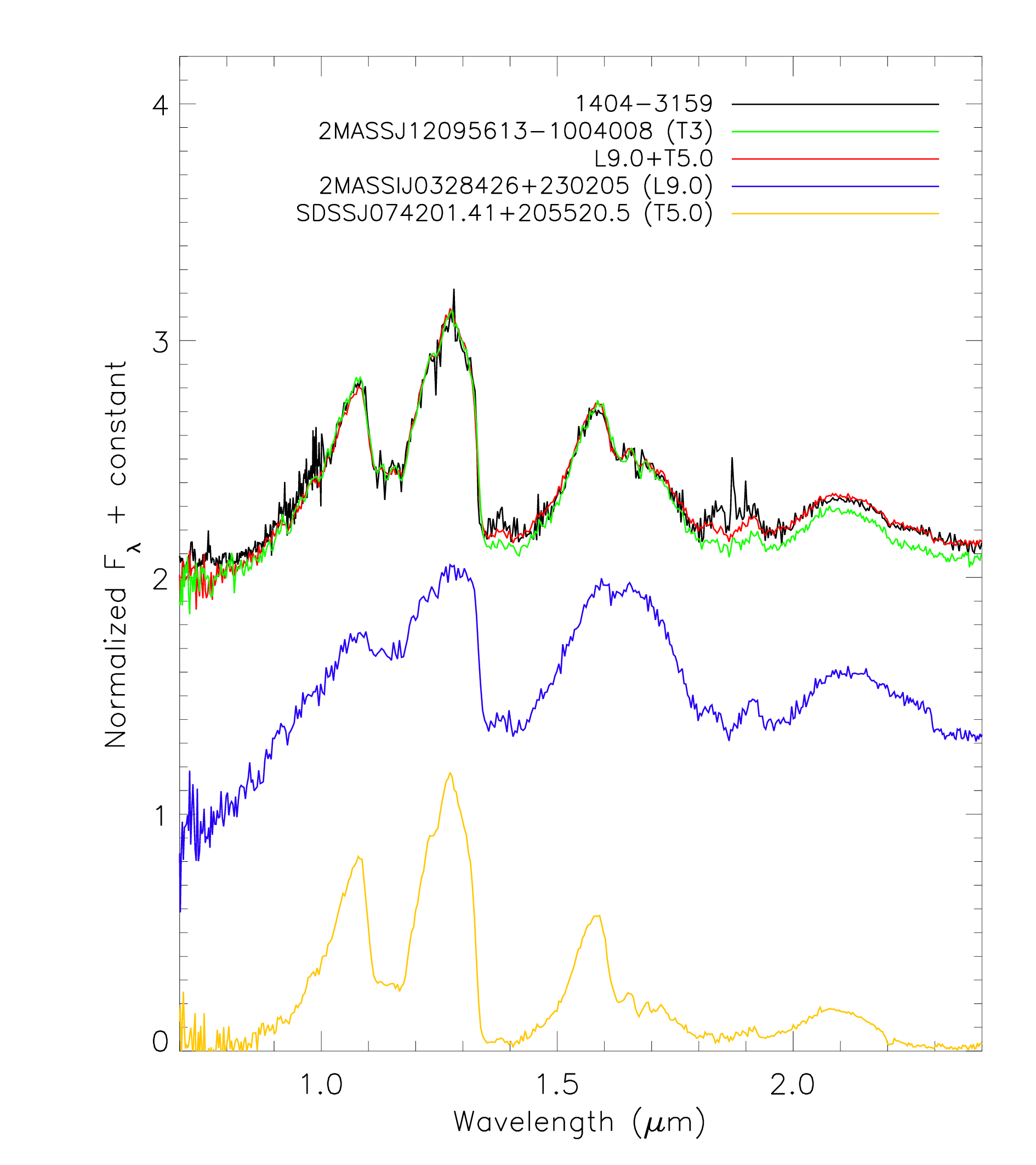}
\caption{The spectral deconvolution of the unresolved binaries 0357-4417 (top panel) and 1404-3159 (bottom panel). On each panel we plot the observed spectrum (in black), the best fit standard template (green), the best fit combined template (red) and its two components (blue and yellow).  \label{deconvolution}}
\end{figure}

We estimated the spectral types of the components of these two systems by fitting their spectra with a set of synthetic unresolved templates. We created the synthetic binaries by combining the spectra taken from the already mentioned SpeX-Prism library. The spectra were normalized to one at 1.28$\mu$m, and then scaled to the appropriate flux level using the $M_{\rm J} - $Spectral type relation presented in \citet{2010A&A...524A..38M}. The results of this fitting are presented in Figure \ref{deconvolution}. For each target we plot the observed spectrum (in black), the best fit standard template (green), the best fit combined template (red) and its two components (blue and yellow). For 1404-3159 we obtain a best fit with a L9+T5 ($\pm$1) template, which is in good agreement with the previous results obtained by \citealt{2008ApJ...685.1183L} (T1+T5), \citealt{2010ApJ...710.1142B} (T0+T5), and \citealt{2012ApJS..201...19D} (L9+T5). For 0357-4417, our deconvolution gives a best fit of L4.5+L5 ($\pm$1). Resolved optical spectroscopy obtained by \citet{2006A&A...456..253M} indicates that the system is likely to be composed of a M9 and an L1. We note however that this object was also identified as a probable young object \citep{2008AJ....136.1290R}. Its NIR spectrum indeed shows peculiarities associated with young ages, especially a triangular shaped H band and an enhancement of the flux in the K band. We therefore conclude that this is a young binary system, and we note that our best fit binary template does not reproduce very well the shape of the H band peak. This is because the spectra we used to create our synthetic binaries are mostly field-aged objects.

\section{Kinematics}
The analysis of the kinematics properties of stars can provide useful insights on their nature. It is well known that different populations of stars (i.e. thin disk, thick disk, and halo members) have different velocity distributions in the U,V,W parameter space. Determining the three components of the galactic velocity of our targets can therefore lead to the determination of their membership. To do this we follow the approach of \citet{2003A&A...410..527B}. In this contribution the authors find that the distribution of the three different star populations in the solar neighborhood are defined by a three-dimensional gaussian:

\begin{equation}
\scalebox{0.8}{
$f(U,V,W)= k \cdot exp \left( -\frac{U^2}{2\sigma_U^2} -\frac{(V - V_{asym})^2}{2\sigma_V^2} - \frac{W^2}{2\sigma_W^2} \right)$}
\end{equation}

where

\begin{equation}
k = \frac{1}{(2\pi)^{3/2} \sigma_U \sigma_V \sigma_W}
\end{equation}

and ($\sigma_U,\sigma_V,\sigma_W$) are the characteristic velocity dispersions, $V_\textrm{asym}$ is the velocity lag for each component behind the galactic rotation.

So if we determine U, V and W for each target, the probability to belong to one of the components (e.g. the thin disk) is given by:

\begin{equation}
P_{Tn} = \frac{X_{Tn} f_{Tn}}{X_{Tn} f_{Tn} + X_{Tk} f_{Tk} + X_{H} f_{H}}
\end{equation}

where $f_{Tn}$, $f_{Tk}$, $f_{H}$ are the velocity distribution $f(U,V,W)$ for thin disk, thick disk, and halo respectively, and X$_{Tn}$, X$_{Tk}$, X$_{H}$ are the observed fraction of objects of each component. The values adopted for $X, \sigma_U, \sigma_V, \sigma_W$ and $V_\textrm{asym}$ for each component are those listed in \citet{2003A&A...410..527B}.

However, to determine the components of the galactic velocity of our objects, we need the radial velocity of the dwarfs. Given that none of our targets has radial velocity measurements, to compute the membership probabilities we follow this approach: first we assume that our objects follow the radial velocity distribution of brown dwarf in the solar neighborhood, which is a gaussian profile centered on 0 km s$^{-1}$ with a sigma of 34 km s$^{-1}$ \citep[e.g.][]{2010AJ....139.1808S}. Then for each target we assume 10000 radial velocity values randomly taken from the gaussian distribution, and for each of these values we calculate a value for $P_{Tn}$, $P_{Tk}$, and $P_{H}$. The uncertainties on the parallax and proper motion are treated the same way, so drawing 10000 values from a gaussian distribution centered on the values given in Table \ref{astro}, and with the associated sigma. Finally, we assume the average value as the membership probability of each target.

The probabilities obtained are listed in Table \ref{prob}. As we can see, all of our targets are disk members ($P_{Tn} + P_{Tk} >$ 99$\%$).

\begin{deluxetable}{l c c c}
\tablewidth{0pt}
\tablecaption{The membership probability for our targets. \label{prob}}
\tablehead{Object & $P_{\rm Tn}$ & $P_{\rm Tk}$ & $P_{\rm H}$ }
\startdata
0032-4405 &  85 & 15 & 0 \\
0058-0651 &  90 & 10 & 0 \\
0109-5100 &  90 & 10 & 0 \\
0147-4954 &  77 & 23 & 0 \\
0219-1939 &  89 & 11 & 0 \\
0230-0953 &  90 & 10 & 0 \\
0239-1735 &  89 & 11 & 0 \\
0257-3105 &  88 & 12 & 0 \\
0539-0059 &  86 & 14 & 0 \\
0614-2019 &  85 & 15 & 0 \\
0719-5051B & 90 & 10 & 0 \\
0928-1603 &  84 & 16 & 0 \\
1246-3139 &  87 & 13 & 0 \\
1331-0116 &  62 & 38 & 0 \\
1404-3159 &  88 & 12 & 0 \\
1753-6559 &  85 & 15 & 0 \\
1936-5502 &  83 & 17 & 0 \\
2045-6332 &  86 & 14 & 0 \\
2209-2711 &  88 & 12 & 0 \\
2310-1759 &  86 & 13 & 0 \\
2346-5928 &  73 & 27 & 0 \\
\enddata
\tablecomments{$P_{\rm Th}$, $P_{\rm Tk}$, and $P_{\rm H}$ are the probabilities of a brown dwarf being a thin disk, thick disk, or a halo object respectively.}
\end{deluxetable}

We note that the sample of brown dwarfs studied in \citet{2010AJ....139.1808S} is formed mostly of thin disk objects (90$\%$). The sigma of the distribution derived in that paper is therefore dictated by the thin disk dwarfs, and it can introduce a bias in the results presented in our Table \ref{prob}. So we tested the membership assigned with our simulation using other two purely kinematic methods. 

\begin{figure}
\includegraphics[width=7.5cm, height=7.3cm]{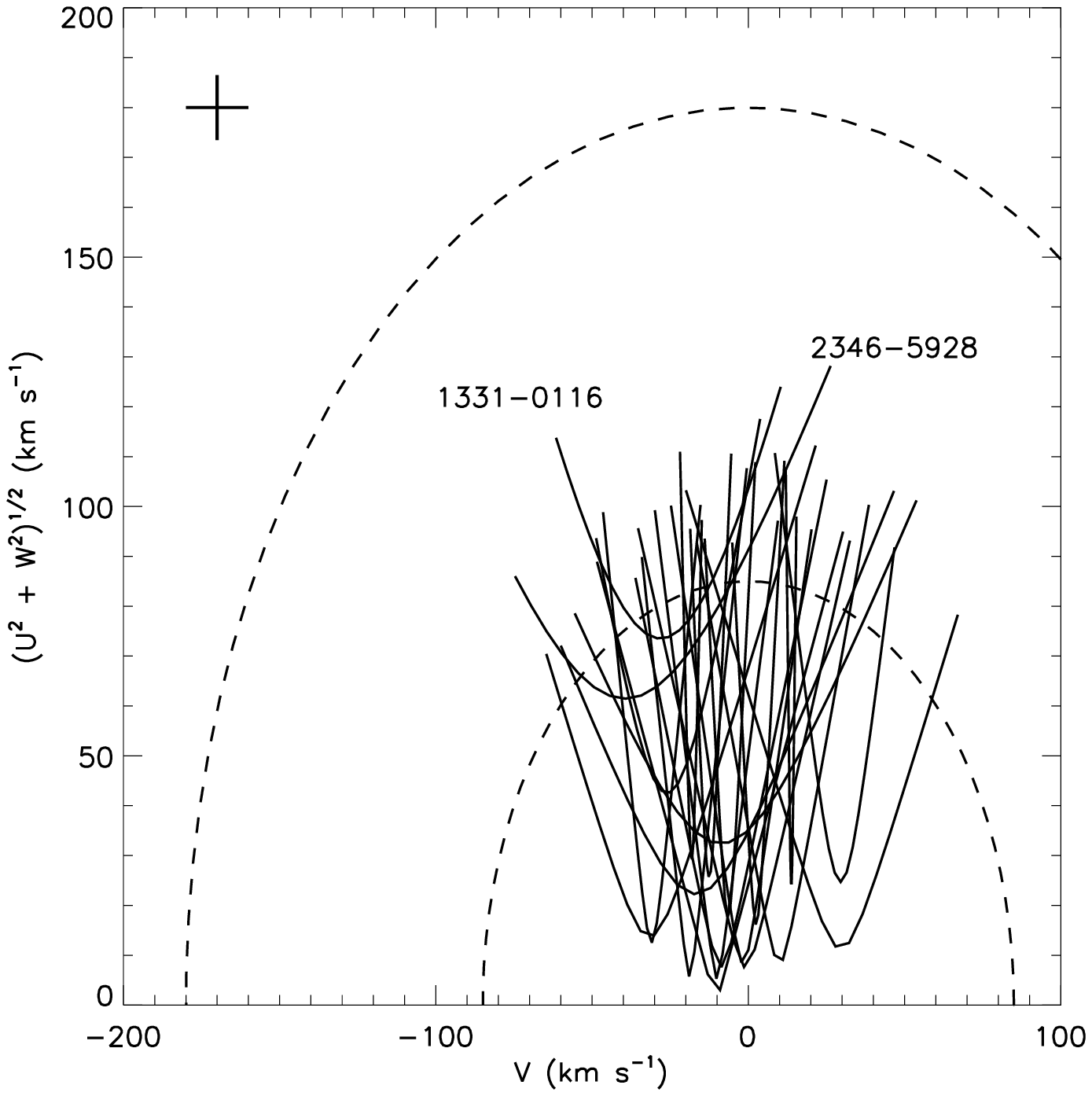}
\includegraphics[width=7.5cm, height=7.3cm]{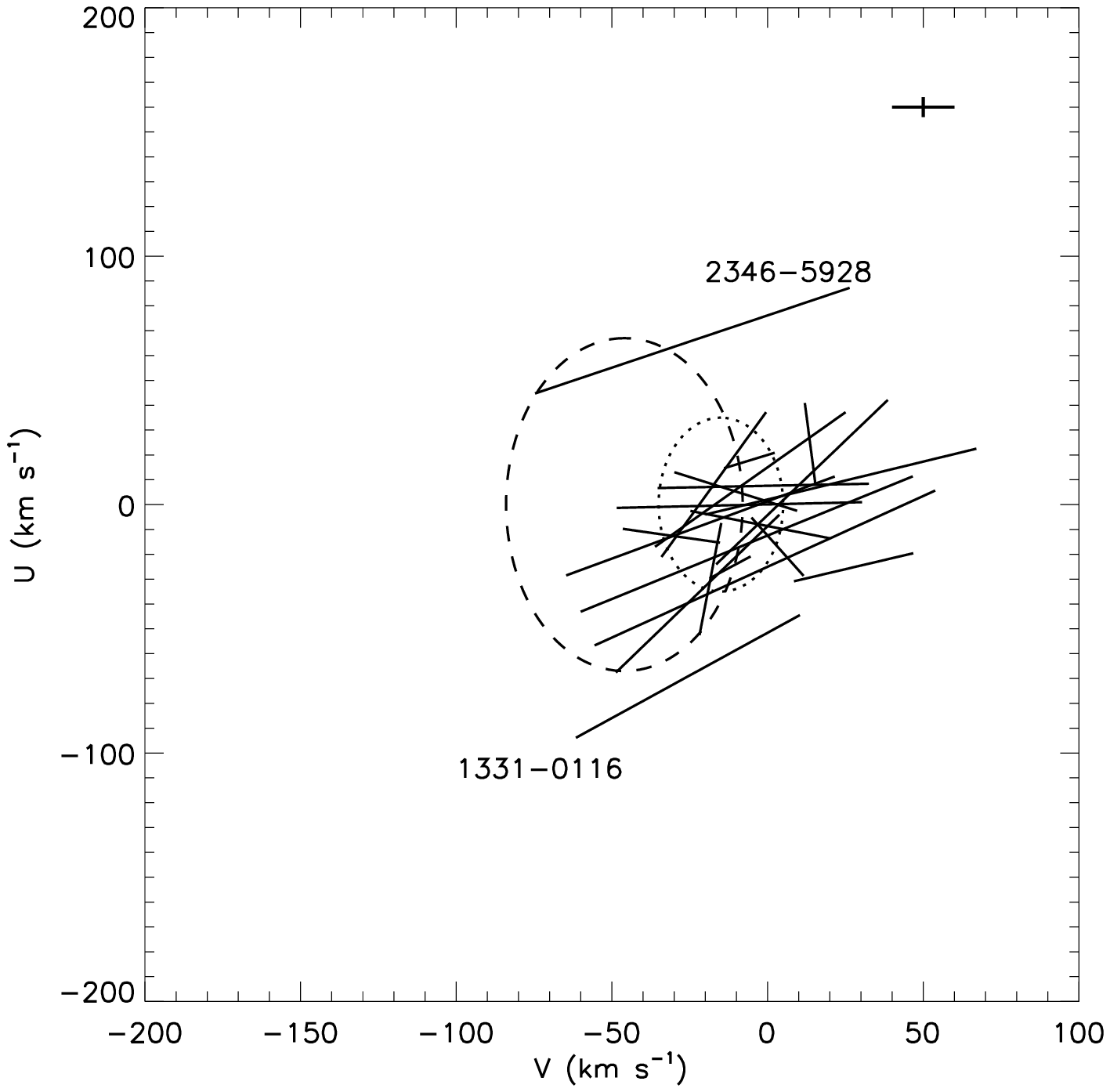}
\caption{The kinematics of the sample. \textit{Top panel}: Toomre diagram. The solid lines represent the velocity ranges of each brown dwarf, obtained assuming a radial velocity range of -100/+100 km s$^{-1}$. The dashed circles are the boundary between thin disk and thick disk stars (inner circle) and between thick disk and halo stars (outer circle), as used by \citet{2004oee..symp..154N}. \textit{Bottom panel}: $U-V$ plot showing the velocity ranges obtained for our targets. Overplotted for reference are the velocity ellipsoids obtained by \citet{2003A&A...410..527B} for thin disk (dotted) and thick disk (dashed) respectively. In each plot the outliers are labelled. Typical uncertainties are shown in the top-left and top-right corner, respectively. \label{toomre}}
\end{figure}

One is the classical Toomre diagram as used by \citet{2004oee..symp..154N} to discriminate between thin disk, thick disk and halo stars. For each of our targets we used the parallaxes and proper motions presented here and we calculated a range for their $UVW$ velocities, assuming that their radial velocities are in a conservative range of -100/+100 km s$^{-1}$. The results are shown in the top panel of Figure \ref{toomre}. The $UVW$ ranges obtained result in an almost parabolic curve for each target. The dashed circles represent the boundaries between thin disk and thick disk stars (inner circle) and between thick disk and halo stars (outer circle). All except two of the targets fall mostly into the thin disk selection area, a result which is consistent with the high thin disk probability derived with the previous method. The two exceptions are 1331-0116 and 2346-5928 which velocity ranges fall mostly into the thick disk selection area, consistently with their slightly higher probability of being thick disk objects ($P_{\rm Tk}$ = 38$\%$ and 27$\%$ respectively). We note, however, that 0147-4954 despite having similar $P_{\rm Tk}$ is instead among the rest of the sample, with its velocity ranges falling mostly into the thin disk selection area.

The second method we used is a direct comparison of the $UVW$ velocity ranges obtained with the velocity ellipsoids defined in \citet{2003A&A...410..527B}. The results are shown in the bottom panel of Figure \ref{toomre}. The memberships assigned based on this criteria are consistent with those obtained by the Toomre diagram, with 2346-5928 falling mostly inside the thick disk ellipsoid, and 1331-0116 falling just outside of the thick disk ellipsoid, but being consistent with a thick disk membership when we consider the uncertainties on the proper motion and parallax. 

We also used the kinematics information to check for the possible membership of our targets to one of the known young moving groups (hereafter MG). In order to do this we followed the method described in \citet{2010MNRAS.402..575C}, that we summarize here.

We considered five MGs: the Pleiades, Castor, Hyades, Sirius (also known as Ursa Major) and IC2391. For each of our targets, using the measured proper motions, we calculated the corresponding proper motion towards the convergent point of each moving group ($\mu_{\rm tcp}$) and the proper motion perpendicular to that direction ($\mu_{\rm pcp}$) using the equations derived by \citet{1992MNRAS.257..257R}. For each MG, we allowed for a scatter in velocity of $\pm$5 km s$^{-1}$ to take into account the intrinsic scatter of the MG and the additional scatter due to gravitational interaction of the MG members with disk stars (disk heating). We converted the velocity scatter into a proper motion scatter using our measured parallaxes. Finally, an object was considered as a MG candidate member if its $\mu_{\rm pcp}$ was less than the estimated scatter or if its 1$\sigma$ interval overlapped with the scatter. Given that we do not have any radial velocity measurement for our targets, we can only classify them as candidate members.

Eleven of the objects presented here are candidate members of at least one of the MGs considered. For each of them we used the $UVW$ velocities determined above (i.e. assuming a $V_{\rm rad}$ in the -100/+100 km s$^{-1}$ range), and applied the selection criteria defined in \citealt{2010MNRAS.402..575C} (see their Figures 5 and 6). This allowed us to further assess the membership of the candidates, and also to derive a $V_{\rm rad}$ range for which the objects would be a member of the MG. Two of our targets passed this second selection: 0032-4405 and 2209-2711. They are both candidate members of the Pleiades and they require their $V_{\rm rad}$ to be in the range 5$-$25 and 10$-$30 km s$^{-1}$ respectively to be members of the MG. We will discuss further their properties in Section \ref{individual}.

\section{Physical parameters}
The knowledge of the distances to our objects allows us to further investigate their nature, determining their physical properties such as bolometric luminosity ($L_{\rm bol}$) and effective temperature ($T_{\rm eff}$).

We determined the effective temperature via model fitting of the observed spectra, using the new version of the atmospheric models presented in \citet[hereafter BT-Settl]{2011ASPC..448...91A}. We followed three different approaches. 

One approach (hereafter method 1) is to scale each model to match the observed flux using the geometric scaling factor, given by the ratio of the distance over the radius of the object squared. We do not know the radii of our targets, but we can assume they all have R = 1.0$\pm$0.2 $R_{\rm Jup}$. The evolutionary models in fact show that brown dwarfs tend to contract quite quickly ($\sim$500 Myr) and reach similar final radii, independent of their mass \citep[e.g.][]{1998A&A...337..403B}. We then determined the best fit model via $\chi^2$ fitting. This method makes use of the astrometric information, but relies on a strong assumption on the radius. This can introduce a bias especially for young objects, which radii can be systematically larger than the assumed one. The random error on the temperature is in most cases dominated by the uncertainty on the radius.

The second approach (hereafter method 2) does not use the parallax but scales each model using the measured infrared photometry (2MASS JHK$_s$ and $WISE$ W1-W2-W3-W4) and then determines the best fit model via $\chi^2$ fitting. The used scaling factor is the median of the seven values given by the magnitudes. In this case we do not rely on any assumption regarding the radii of the targets, but the use of the photometry can introduce other biases. Mostly, in the case of unresolved binaries, the photometric scaling factor would bias the derived temperature towards higher values. The random error introduced by the uncertainty on the photometric values are negligible compared to our floor precision level, which is dictated by the model grid spacing.

The last method we adopted (hereafter method 3) is to normalize both the models and the measured spectrum to 1 at 1.28 $\mu$m and then perform the $\chi^2$ fitting.  This method does not rely on any assumptions on the radii of the object, and it is not prone to any systematic introduced by the photometry. The only constraint on the final temperature is given by the shape of the object's spectrum. However with this approach gravity and metallicty of the dwarf are important parameters of the fit. Given the known degeneracy between the two \citep[e.g.][]{2005ARA&A..43..195K} their determination is very uncertain, and can bias the temperature we obtain especially for the peculiar objects.

We assume as our final value the weighted average of the three values, as this approach minimizes the systematic errors. We then calculate the bolometric luminosity of the targets following the Stefan-Boltzmann law:

\begin{equation}
L_{\rm bol} = 4\pi \sigma R^2 T_{\rm eff}^4
\end{equation}

The results are presented in Table \ref{teff}. In the first column we indicate the target short name, in the second one its spectral type, in the third the expected temperature according to the temperature-spectral type relation given by \citet{2009ApJ...702..154S}, in the fourth the derived temperature and in the last column the bolometric luminosity. We note that hottest objects tend to have systematically higher uncertainties on $T_{\rm eff}$ compared to the colder ones. This is probably a consequence of the fact that at hotter temperatures the contribution of the optical part of the spectrum becomes significant, hence our fit based solely on the near-infrared portion of the spectrum becomes less and less accurate. Therefore the scatter between the three methods increases.

\begin{figure*}
\centerline{
\includegraphics[width=8cm]{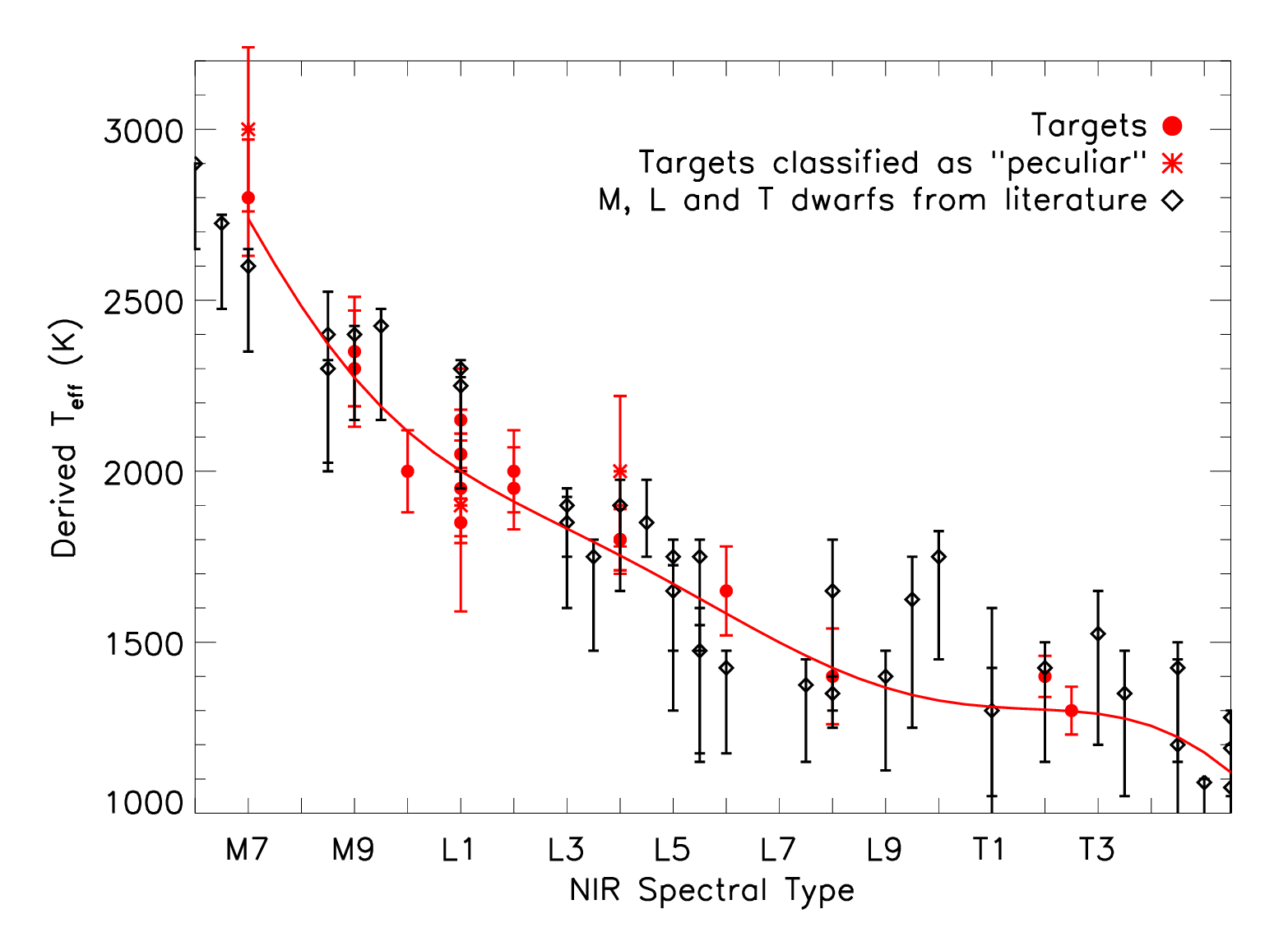}
\includegraphics[width=8cm]{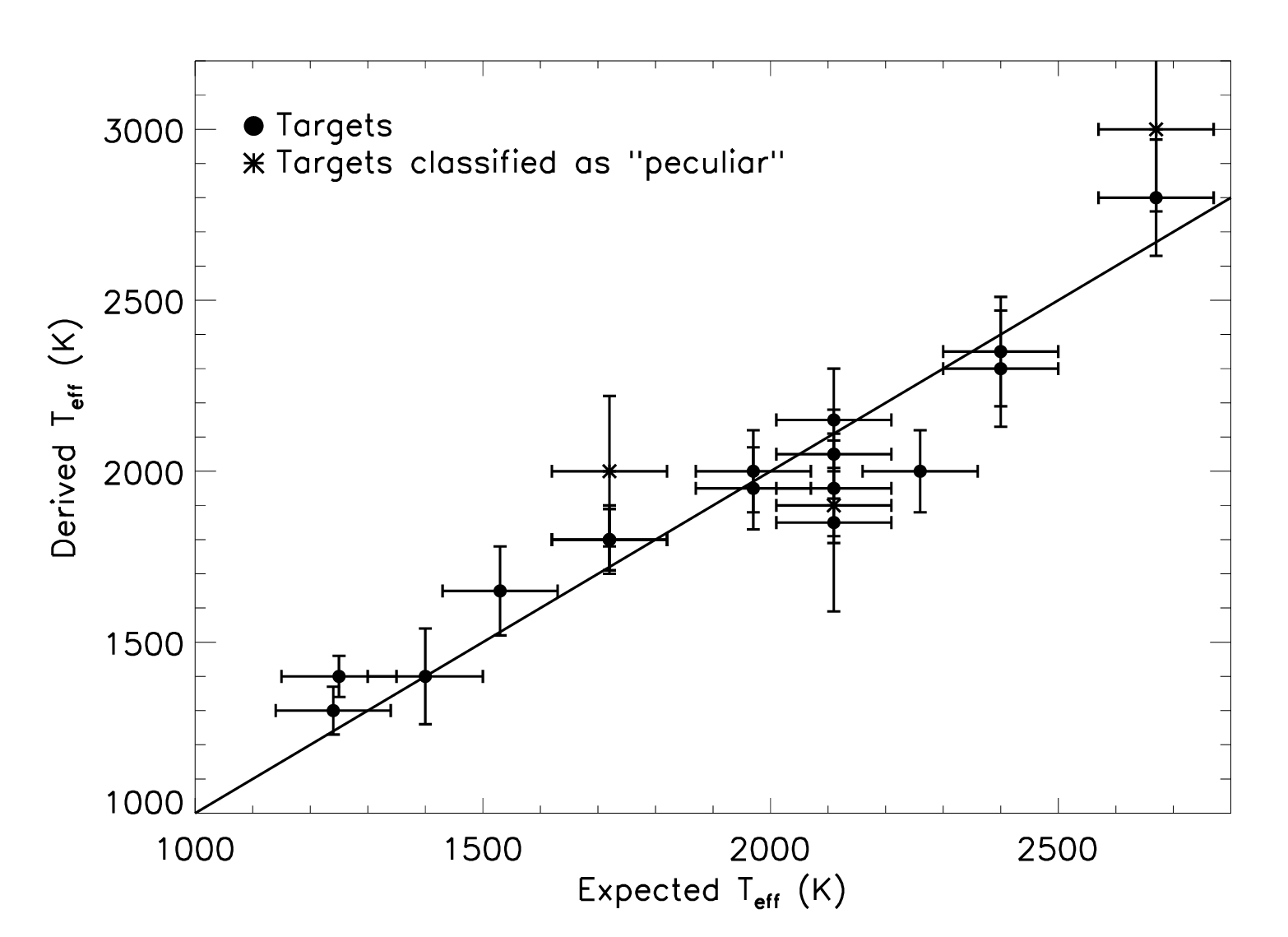}
}
\caption{\textit{Left panel:} The effective temperature of our targets (filled circles) as a function of their spectral types. Peculiar objects are plotted as asterisks. Overplotted as diamonds are objects taken from \citet{2004AJ....127.3516G} and \citet{2010A&A...524A..38M}. The red line is our $7th$ order polynomial fit to the M7 to T8 sequence, excluding the peculiar objects. \textit{Right panel:} A comparison between the $T_{\rm eff}$ derived in this paper and those predicted using the polynomial relation from \citet{2009ApJ...702..154S}. The solid line is the bisector of the plot. Although generally consistent with each other, our derived temperatures are systematically higher than the expected ones. \label{teff_plot}}
\end{figure*}

\begin{figure}
\centerline{
\includegraphics[width=7.5cm]{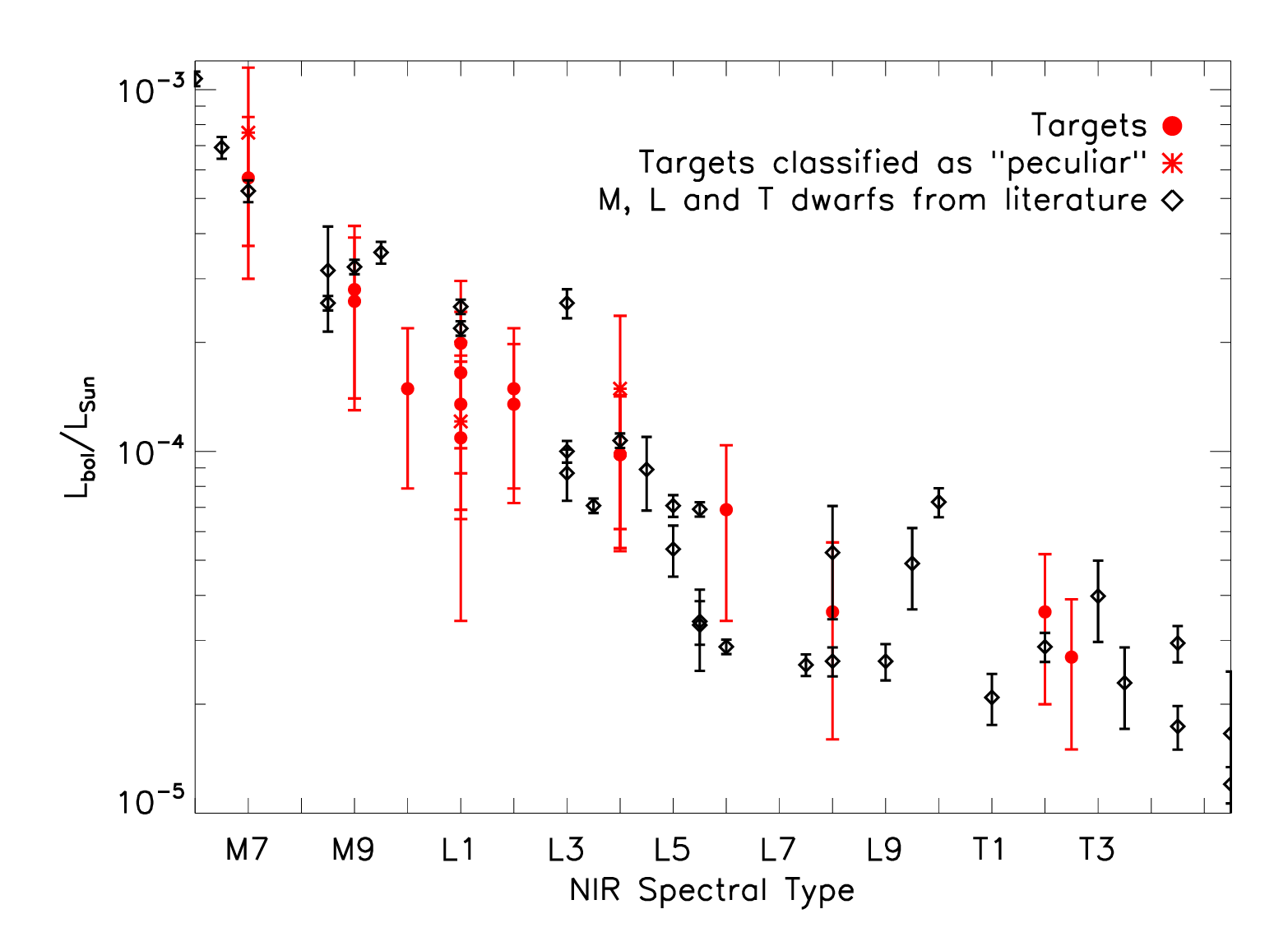}
}
\caption{The bolometric luminosity of our targets (filled circles) as a function of their spectral types. Peculiar objects are plotted as asterisks. Overplotted as diamonds are objects taken from \citet{2004AJ....127.3516G} and \citet{2010A&A...524A..38M}. \label{lbol}}
\end{figure}

The results are also plotted in Figure \ref{teff_plot} and \ref{lbol}. In the left panel of Figure \ref{teff_plot} we show the $T_{\rm eff}$ we derived here for our targets (plotted as filled circles) as a function of the spectral type. Objects that were classified as ``peculiar'' are marked as asterisks. Overplotted as diamonds are objects taken from \citet{2004AJ....127.3516G} and \citet{2010A&A...524A..38M}. The red line is our seventh-order polynomial fit to the sequence for spectral types from M7 to T8, excluding the peculiar objects. The polynomial obtained is:

\begin{align}
&T_{\rm eff} = &-&\, 1613.82 + 3561.47\,{\rm SpT} - 975.953\,{\rm SpT^2} \nonumber\\ \vspace{0.5mm}
&&+&\, 129.141\,{\rm SpT^3} - 9.46896\,{\rm SpT^4} \nonumber\\ \vspace{0.5mm}
&&+&\, 0.390319\,{\rm SpT^5} - 0.00843736\,{\rm SpT^6} \nonumber \\ \vspace{0.5mm}
&&+&\, 0.0000742110\,{\rm SpT^7} \quad (\pm 140)\quad {\rm K}
\end{align}

We have chosen a 7th order polynomial as it is the one that minimizes the reduced $\chi^2$ and the uncertainties of the individual coefficients.

Our new fit suggests a change in the slope of the sequence at the transition between the M and L dwarfs. This may be an effect of dust formation and its migration into the photosphere, that causes a more rapid evolution of the spectral features as a function of $T_{\rm eff}$. The transition from M to L spectral types is indeed characterized by the formation of aluminum-, calcium- and titanium-bearing molecules such as perovskite (CaTiO$_3$), corundum (Al$_2$O$_3$), and grossite (CaAl$_4$O$_7$), which remove those elements from the atmosphere of the dwarfs. At slightly lower temperature other condensates, like forsterite (Mg$_2$SiO$_4$), enstatite (MgSiO$_3$), and vanadium dioxide (VO$_2$), remove the VO and Si from the atmosphere, causing the alkali metals (Na and K primarily) and the metal hydrides (in particular FeH and CrH) to be the main absorbers in the atmospheres of L dwarfs \citep[see][and references therein for a more detailed description of the chemistry of ultracool atmospheres]{2005ARA&A..43..195K}.

Also, in the L-T transition the sequence is almost flat. This is a known phenomenon, and it is the effect of the onset of the dust settling and of the Collision Induced Absorption (CIA) of the H$_2$ \citep[e.g.][]{2006ApJ...640.1063B,2008ApJ...689.1327S,2011ASPC..448...91A}.

In the right panel of Figure \ref{teff_plot} we present a comparison between the $T_{\rm eff}$ derived in this paper and those predicted by the polynomial relation presented by \citet{2009ApJ...702..154S}. The values are generally consistent with each other, but we note that our estimated temperatures are systematically slightly higher than the predicted ones. The polynomial fit by \citet{2009ApJ...702..154S} is based essentially on the $T_{\rm eff}$ derived in \citet{2004AJ....127.3516G}. In that contribution the authors estimated the bolometric flux using the measured NIR spectra (covering the 0.8$-$2.5 $\mu$m range) and applying a bolometric correction based on the $L'$ photometry only, interpolating between the K and $L'$ band and assuming a Rayleigh-Jeans tail longward of $L'$. This approximation could have led to a systematic underestimation of the bolometric flux, hence of the $T_{\rm eff}$ which would explain the discrepancy in Figure \ref{teff_plot}.

Figure \ref{lbol} shows the bolometric luminosity as a function of the spectral type. Colours and symbols follow the same convention of Figure \ref{teff_plot} and literature objects are the same shown in that Figure as well. The bolometric luminosity decreases smoothly from late-M type object to mid and late-L dwarfs, and from mid-T down to late-Ts. In the L/T transition the luminosity is almost constant, and despite the sparse population (only 8 objects between L7 and T1) we note a high scatter, with difference of a factor of 3-4 between objects of the same or very near spectral type. This scatter is not unexpected, as the L/T transition is known to be populated by a high fraction of unresolved binaries \citep{2013AN....334...32B}, and these objects would clearly result as overluminous compared to single dwarfs of the same spectral type. 

\begin{deluxetable}{c c c c c c c c}
\tablewidth{0pt}
\tablecaption{Luminosity and effective temperature of the targets. \label{teff}}
\tablehead{Object & Sp. & Exp. $T_{\rm eff}$ & $T_{\rm eff}$ & $T_{\rm eff}$ & $T_{\rm eff}$ & Calc. $T_{\rm eff}$ & $L_{\rm bol}$ \\
short name & Type & (K) & method 1 & method 2 & method 3 & (K) & ($L_\odot$) }
\startdata
0032-4405 & L4 pec & 1720 & 2100$\pm$360 & 2000$\pm$100 & 2000$\pm$100 & 2000$\pm$220 & 1.49$\pm$0.88$\times$10$^{-4}$ \\
0058-0651 & L1     & 2110 & 2250$\pm$170 & 2000$\pm$110 & 2000$\pm$110 & 2050$\pm$130 & 1.65$\pm$0.78$\times$10$^{-4}$ \\
0109-5100 & M7     & 2670 & 2800$\pm$160 & 2800$\pm$170 & 2800$\pm$170 & 2800$\pm$170 & 5.7$\pm$2.7$\times$10$^{-4}$ \\
0147-4954 & M9     & 2400 & 2700$\pm$200 & 2000$\pm$120 & 2800$\pm$160 & 2350$\pm$160 & 2.8$\pm$1.4$\times$10$^{-4}$ \\
0219-1939 & L1     & 2110 & 2650$\pm$200 & 1900$\pm$90  & 1900$\pm$90  & 1950$\pm$140 & 1.35$\pm$0.66$\times$10$^{-4}$ \\
0230-0953 & L6     & 1530 & 1500$\pm$110 & 1700$\pm$170 & 1800$\pm$100 & 1650$\pm$130 & 6.9$\pm$3.5$\times$10$^{-5}$ \\
0239-1735 & M9     & 2400 & 2250$\pm$200 & 2100$\pm$120 & 2800$\pm$180 & 2300$\pm$170 & 2.6$\pm$1.3$\times$10$^{-4}$ \\
0257-3105 & L8     & 1400 & 1500$\pm$90  & 1700$\pm$210 & 1300$\pm$80  & 1400$\pm$140 & 3.6$\pm$2.0$\times$10$^{-5}$ \\
0539-0059 & L4     & 1720 & 1700$\pm$90  & 1800$\pm$90  & 2000$\pm$100 & 1800$\pm$90  & 9.8$\pm$4.4$\times$10$^{-5}$ \\
0614-2019 & L2     & 1970 & 2000$\pm$130 & 2000$\pm$120 & 2000$\pm$120 & 2000$\pm$120 & 1.49$\pm$0.70$\times$10$^{-4}$ \\
0719-5051 & L0     & 2260 & 2550$\pm$150 & 1900$\pm$100 & 1900$\pm$100 & 2000$\pm$120 & 1.49$\pm$0.70$\times$10$^{-4}$ \\
0928-1603 & L2     & 1970 & 2000$\pm$150 & 1900$\pm$110 & 2000$\pm$110 & 1950$\pm$120 & 1.35$\pm$0.63$\times$10$^{-4}$ \\
1246-3139 & T2     & 1250 & 1350$\pm$70  & 1400$\pm$60  & 1400$\pm$60  & 1400$\pm$60  & 3.6$\pm$1.6$\times$10$^{-5}$ \\
1331-0116 & L1 pec & 2110 & 1350$\pm$140 & 1900$\pm$80  & 2200$\pm$100 & 1900$\pm$110 & 1.21$\pm$0.56$\times$10$^{-4}$ \\
1753-6559 & L4     & 1720 & 1800$\pm$110 & 1800$\pm$80  & 1900$\pm$90  & 1800$\pm$90  & 9.8$\pm$4.4$\times$10$^{-5}$ \\
1936-5502 & L4     & 1720 & 1800$\pm$120 & 1900$\pm$90  & 1800$\pm$80  & 1800$\pm$100 & 9.8$\pm$4.5$\times$10$^{-5}$ \\
2045-6332 & L1     & 2110 & 3100$\pm$220 & 2200$\pm$100 & 1900$\pm$100 & 2150$\pm$150 & 1.99$\pm$0.97$\times$10$^{-4}$ \\
2209-2711 & T2.5   & 1240 & 1100$\pm$70  & 1400$\pm$70  & 1400$\pm$60  & 1300$\pm$70  & 2.7$\pm$1.2$\times$10$^{-5}$ \\
2310-1759 & L1     & 2110 & 3100$\pm$430 & 1800$\pm$80  & 1800$\pm$100 & 1850$\pm$260 & 1.09$\pm$0.75$\times$10$^{-4}$ \\
2346-5928 & M7 pec & 2670 & 3300$\pm$350 & 2700$\pm$150 & 3300$\pm$180 & 3000$\pm$240 & 7.6$\pm$3.9$\times$10$^{-4}$ \\
\enddata
\tablecomments{For each object we list the NIR spectral type derived in this paper, the expected temperature according to \citet{2009ApJ...702..154S} polynomial relation, the calculated temperature using the three methods described in the text, the final value adopted, and the bolometric luminosity. The uncertainty on the expected temperatures is $\pm$100 K.}
\end{deluxetable}

\section{Comments on Individual Objects \label{individual}}

\subsection{EROS-MP~J0032$-$4405 (0032-4405)}
This object was identified as a possible young object by \citet{2008AJ....136.1290R}. The NIR spectrum shows indeed a slightly triangular-shaped H band, which is associated with low gravity \citep[hence with young age, e.g.][]{2001MNRAS.326..695L,2006ApJ...639.1120K}. It is also the brightest L4 in our sample ($M_{\rm J} = 11.45, M_{\rm H} = 10.53, M_{\rm K} = 9.94$), a fact that would be consistent with a young nature, as 0032-4405 would not have fully contracted to its final radius. With a larger radius, the object would then look slightly brighter than its spectral analogue of older age. We therefore classify this object as L4 pec. We have also identified this object as a possible member of the Pleiades, a fact that would be in agreement with its young age.

\subsection{SSSPM~J0109$-$5100 (0109-5100)}
Our derived NIR spectral type (M7) differs by five subtypes from the one published in \citealt{2005A&A...440.1061L} (L2), while it is consistent within the uncertainties with the optical classification derived in the same paper (M8.5). No clear signs of peculiarity are present in the spectrum of 0109-5100 that could justify this discrepancy. However as noted by \citet{2005A&A...440.1061L} their NIR classification, based on the spectral indices defined in \citet{1999AJ....117.1010T}, \citet{2000vlms.conf..119M}, and \citet{2001AJ....121.1710R}, is systematically offset towards later types when compared to their optical classification, and with a large scatter. The authors therefore assigned to the object a type of M8.5 based on the optical spectrum only. Moreover, our new NIR classification is based on a different system, which is the direct comparison of our spectra to the new standard templates defined in \citet{2010ApJS..190..100K}.

\subsection{DENIS-P~J035726.9$-$441730 (0357-4417)}
This target is a known unresolved binary, identified by \citet{2003AJ....126.1526B}. As we discussed in Section \ref{unres_bins}, the spectrum of this object shows signs of low-gravity, which is associated with young ages. We assign an unresolved spectral type of L2 pec, because the L2 standard template is the one that gives the best fit in the J band. The spectral deconvolution gives spectral types for the individual components of the system of L4.5 and L5, which are much later than those derived via resolved optical spectroscopy by \citealt{2006A&A...456..253M} (M9+L1). The discrepancy is probably due to the fact that the templates we employed for the deconvolution are ``normal'' field M and L dwarfs, thus they do not reproduce well the H and K band peculiarities typical of young dwarfs.

\subsection{2MASS~J07193188$-$5051410 (0719-5051)}
This object forms a common proper motion pair with 0719-5050, as already noted in AHA11. We obtained an infrared spectrum for both objects. We confirm the spectral classification of L0 for 0719-5051, as obtained by \citet{2008AJ....136.1290R}. For the companion, we derive a spectral type of M4, based on the spectral fitting with the templates obtained from the IRTF spectral library, which is consistent with the photometric estimate of AHA11.

Given the relatively limited time-span of our observations, it is impossible to detect hints of orbital motion for the system. The predicted average astrometric acceleration terms along the X and Y axis \citep[e.g.][]{1999PASP..111..169T}, assuming masses of 0.1 $M_\odot$ and 0.08 $M_\odot$ for the two components of the system, given the projected separation and distance (and averaging over all other orbital parameters) are well below 1 $\mu$as yr$^{-1}$.

\subsection{2MASSW~J1004392$-$333518 (1004-3335)}
This object is in a common proper motion system with LHS~5166 (AHA11). the infrared spectrum obtained for 1004-3335 indicates a spectral type of L5, in good agreement with the optical spectral type of L4 obtained by \citet{2002ApJ...575..484G}. For the bright companion, LHS~5166, we derive a spectral type of M4, slightly later than the M3 found by AHA11, but in agreement with the dM4.5e published in \citet{2005AN....326..974S}. For the same reasons listed in the previous subsection, it is impossible to detect any hint of orbital motion for this system.

\subsection{SDSS~J133148.92$-$011651.4 (1331-0116) \label{sdss1331}}
\begin{figure*}
\includegraphics[width=16cm]{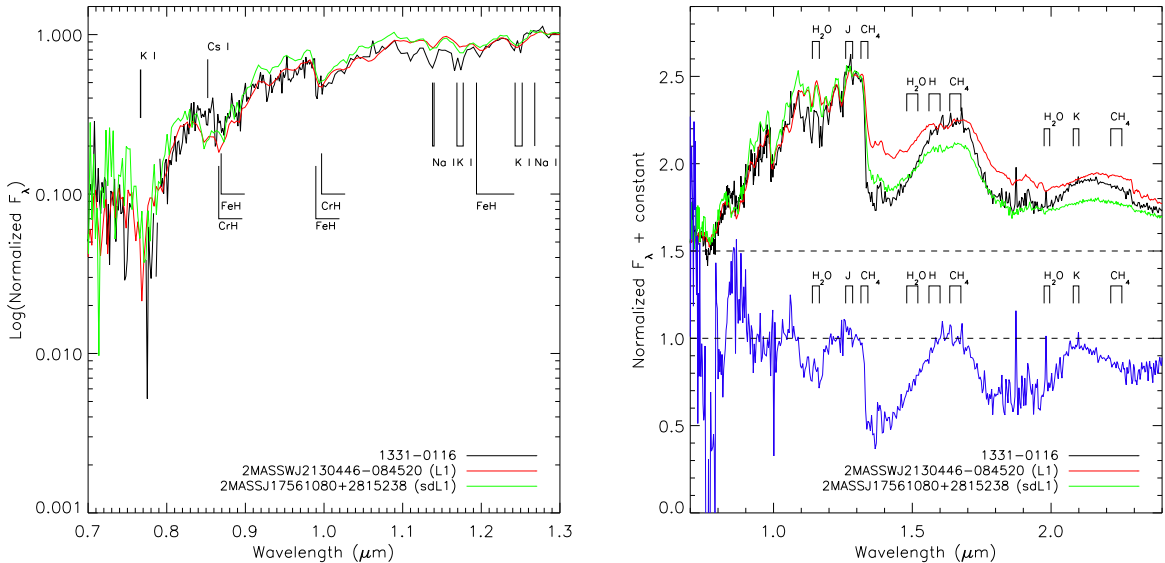}
\caption{The spectrum of 1331-0116. \textit{Left}: a zoom to the optical and J-band spectrum; \textit{Right}: the entire spectrum. Overplotted in both panels are the L1 standard 2MASSW~J2130446$-$084520 (red) and the sdL1 2MASS~J17561080+2815238 (green). In blue we show the flux ratio between the target and the L1 spectroscopic standard. Marked with a dashed line are the zero flux level of the normalized spectra and the 1 level of the flux ratio.}
\label{sdss1331_spectrum}
\end{figure*}

The spectrum of this object is presented in Figure \ref{sdss1331_spectrum} with the spectrum of the L1 standard 2MASSW~J2130446$-$084520 and of the sdL1 2MASS~J17561080+2815238 overplotted for comparison, in red and green respectively. The overall slope of the optical spectrum of 1331-0116 is well matched by the L1 standard, the target however shows the peculiar signs of subdwarfs, i.e. stronger absorption by alkali metals, for examples the depth of the K I line at 0.78 $\mu$m and the doublets at 1.169-1.177, and 1.244-1.252 $\mu$m. In the near-infrared range, the L1 standard matches the flux level at the peak of the H band, but we can clearly see stronger H$_2$ CIA and also much deeper H$_2$O bands. When compared the sdL1 template, 1331-0116 shows deeper H$_2$O bands at 1.1 and 1.35 $\mu$m, but a much higher flux level in the H and K band. This object was already noted as peculiar in \citet{2004AJ....127.3553K} and low-metallicity was pointed out as the possible explanation for its peculiarity. The parallax and proper motion obtained for it are $\pi_{\rm abs}$ = 67.3$\pm$12.6 mas, $\mu_{\alpha}$cos$\delta$ = -421.9$\pm$5.7 mas yr$^{-1}$ and $\mu_\delta$ = -1039.0$\pm$5.2 mas yr$^{-1}$. The kinematics of 1331-0116 suggests that this object may pertain to a slightly older population, with a probability of 38$\%$ of being a thick disk object (see Table \ref{prob}). We therefore conclude that this object is a slightly metal-poor L1 dwarf, and we classify it as L1 pec. We note that the previous infrared classification, based on spectral indices, was L8$\pm$2.5 \citep{2004AJ....127.3553K}. This discrepancy is not surprising, as the index-based classification for peculiar L dwarfs is not well established, and the criteria used to classify normal objects can therefore lead to uncertain spectral types. The very deep H$_2$O absorption bands are likely to be the reason of the previous late-type classification. However, the L8 and L9 standards do not match the depth of the water absorption, nor the slope and shape of the optical spectrum. Also, our new classification is consistent with the photometry of the object (J-H = 0.98, J-K = 1.39) which is typical of early-L dwarfs.

\subsection{2MASS~J14044941$-$3159329 (1404-3159)}
This object is an unresolved L/T transition binary. Identified by \citet{2008ApJ...685.1183L} via high resolution imaging with HST, the spectral types of the two components were initially estimated to be T1+T5, then corrected to T0+T5 by \citet{2010ApJ...710.1142B}. More recently, \citet{2012ApJS..201...19D} estimated L9+T5.  Our spectral deconvolution also gives L9+T5. 

The parallax and proper motion derived here ($\pi_{\rm abs}$ = 49.2$\pm$3.4 mas, $\mu_\alpha cos \delta$ = 337.6$\pm$1.9 mas yr$^{-1}$ and $\mu_\delta$ = $-$16.3$\pm$2.4 mas yr$^{-1}$) are consistent with the values found by \citet{2012ApJS..201...19D}, who measured an absolute parallax of 42.1$\pm$1.1 mas and proper motion components $\mu_\alpha cos \delta$ = 344.8$\pm$1.0 mas yr$^{-1}$ and $\mu_\delta$ = $-$10.8$\pm$1.4 mas yr$^{-1}$, except for the $\mu_\alpha cos \delta$ component, but this difference maybe due to our derivation of the parallax which assumes single objects.

\subsection{2MASS~J19285196$-$4356256 (1928-4356)}
We classified this object as L4 pec, as its spectrum appears significantly bluer than the L4 standard 2MASS~J21580457$-$1550098. The standard reproduces well the shape and flux level of the J band spectrum, but at longer wavelengths 1928-4356 emits much less flux, which can be an indication of a stronger H$_2$ absorption due to low metallicity. We therefore conclude that 1928-4356 could be a slightly metal poor object.

\subsection{2MASS~J20115649$-$6201127 (2011-6201) \label{2011-6201}}
\begin{figure*}
\includegraphics[width=16cm]{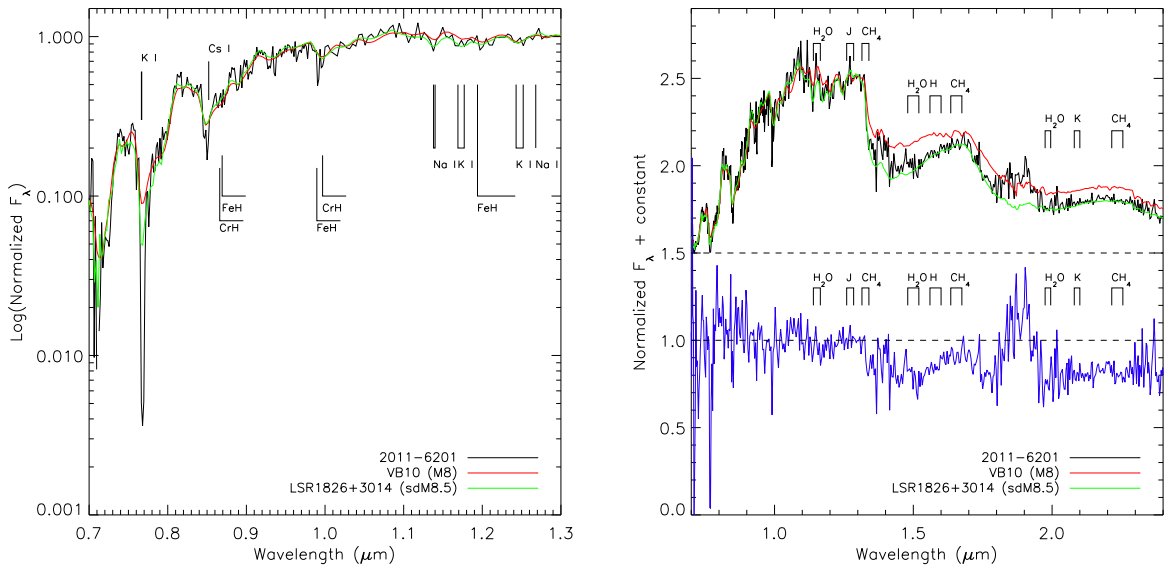}
\caption{Same as Figure \ref{sdss1331_spectrum}, but for 2011-6201. Overplotted in red is the spectrum of the M8 standard VB~10, and in green is the spectrum of the sdM8.5 LSR~1826+3014.} 
\label{2011_spectrum}
\end{figure*}

In Figure \ref{2011_spectrum} we can see that the optical spectrum (left panel) matches quite well the spectrum of the M8 standard VB~10 (overplotted in red), while the NIR spectrum (right panel) shows signs of metal depletion. In particular, we note the flux suppression in the H and K bands and the presence of deeper water absorption bands. These features are associated with low-metallicity and are well matched by the sdM8.5 LSR~1826+3014 (overplotted in green). We calculated the metallicity index $\zeta_{\rm TiO/CaH}$ as defined by \citet{2007ApJ...669.1235L} and found a value of 1.01, which would yield to a classification as a normal dwarf. However, the nature of 2011-6201 is clearly intermediate between a normal dwarf and a subdwarf, and we therefore classify it as a d/sdM8.

\subsection{SIPS2045$-$6332 (2045-6332) \label{2045-6332}}
This is the brightest L1 in our sample ($M_{\rm J} = 10.63, M_{\rm H} = 9.82, M_{\rm K} = 9.22$). The $T_{\rm eff}$ determined via model fitting is also higher than the predicted one. These can be indications of binarity. To investigate further this possibility, we fitted the spectrum of 2045-6332 with our set of unresolved templates. The two components derived by our deconvolution would be L1.0 and T6.0. To assess the significance of this deconvolution we performed an F-test. If $\eta$, which is the ratio of the $\chi^2$ of the two fits (the deconvolution and the one with standard templates) is greater than the critical value $\eta_{\rm crit}$ (which depends on the number of degrees of freedom), than the deconvolution is better than the standard fit with a 99$\%$ significance. In our case, $\eta$ = 1.09, while $\eta_{\rm crit}$ = 1.22. We conclude that the deconvolution is not significant. It still remains possible that the object is an equal (or nearly-equal) spectral type binary. Our deconvolution is not sensitive to these objects, but such kind of binary would clearly appear overluminous and hotter compared to other dwarfs of similar spectral type.

We note also that the H band of the spectrum of 2045-6332 appears slightly triangular, which could be a hint of youth. This can be an alternative explanation to its overluminosity, as young objects have larger radii compared to older, field-aged dwarfs of the same spectral type. 

Further investigation is necessary to determine the nature of this object. In particular, high-resolution imaging is required to address the possibility that this object is a binary system, while optical spectroscopy can help to investigate its young nature.

\subsection{2MASS~J22092183$-$2711329 (2209-2711)}
This is a newly discovered T dwarf. We assign a spectral type of T2.5 as its spectrum shows features which are intermediate between the T2 and the T3 spectral standards (SDSSp J125453.90$-$012247.4 and 2MASS~J12095613 $-$1004008 respectively). This target was also selected as a candidate member of the Pleiades. However, its spectrum does not show any sign of youth. The derived absolute magnitudes and effective temperature are in good agreement with the expected ones.

\subsection{2MASS~J22134491$-$2136079 (2213-2136)}
This object was identified as a low-gravity object by \citet{2009AJ....137.3345C} and classified L0$\gamma$ using its optical spectrum. The NIR spectrum confirms the low-gravity nature of this object. It shows in fact a triangular shaped H band and an enhancement of the flux in the H and K band (compared to a standard template). We classify this object as a L2 pec, as the L2 standard is the one that reproduces better the shape of the J band and the depth of the water absorption band between the J and H band. 

\subsection{SIPS2346$-$5928 (2346-5928)}
\begin{figure}[h]
\includegraphics[width=7.5cm]{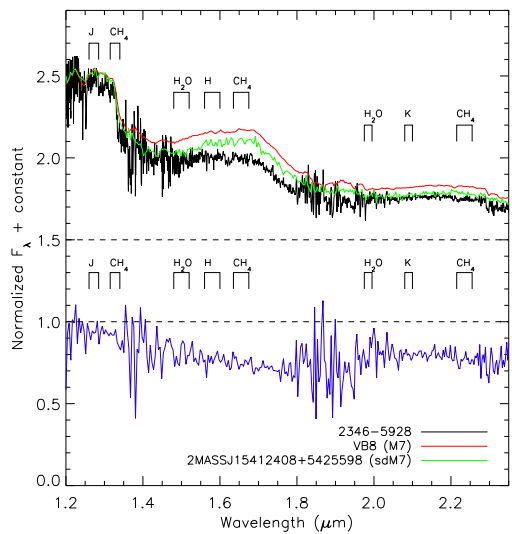}
\caption{The spectrum of 2346-5928. Overplotted in red is the spectrum of the M7 standard VB~8, and in green is the spectrum of the sdM7 2MASS~J15412408+5425598. \label{2346}}
\end{figure}

This newly discovered M7 dwarf appears significantly bluer than the M7 standard VB~8. In Figure \ref{2346} we can see that the sdM7 2MASS J15412408+5425598 reproduces better the depth of the water absorption bands and the flux level in the K band. The H band of 2346-5928 is slightly bluer even when compared to the sdM7. The kinematics suggests that this object could be a member of the galactic thick disk, and we therefore conclude that 2346-5928 is a metal-poor M dwarf. We do not have an optical spectrum for this target, so we cannot apply the criteria defined by \citet{2007ApJ...669.1235L} and therefore we cannot assign a metallicty class. So we decide to classify it as M7 pec. The derived $T_{\rm eff}$ is slightly higher than the prediction, but consistent with the findings for the other M7 of the sample, 0109-5100. The big uncertainty on the $T_{\rm eff}$ is given mainly by the very high temperature (3300 K) that we derive using method 3 (i.e. normalizing the models). This could be due to the peculiarity of 2346-5928, whose blue spectrum is better fitted by a hotter model.

\section{Summary and Conclusions}
We have presented here NIR spectra of 52 M, L and T dwarfs from the PARSEC program, and parallaxes for 21 of them. Ten of these objects are new discoveries. The new parallaxes contribute to further populate the low-mass end of the H-R diagram, especially in the L0-L4 spectral range. The combination of astrometry and spectroscopy has allowed us to investigate in detail the nature of some peculiar objects like the young, low-gravity dwarfs 0032-4405, 0357-4417, and 2213-2136; the unresolved binaries 0357-4417 and 1404-3159; the metal-poor dwarfs 1331-0116, 1928-4356, 2346-5928 and 2011-6201. Also, combining the spectra obtained with photometric data, parallaxes and atmospheric models, we derived effective temperature and bolometric luminosity for 21 of our targets. These new results seem to suggest a change in the slope of the $T_{\rm eff}$ - Spectral type sequence at the M/L spectral type transition. This feature could be due to the formation of dust clouds in the atmospheres of brown dwarfs, and the subsequent migration of the clouds into the photosphere. An increased sample of late-M and early-L with measured $T_{\rm eff}$ will help to constrain better the polynomial relation and understand the physics of the transition.

These 21 objects represent the first sub-sample of parallaxes obtained by PARSEC. The spectroscopic follow-up is in progress, to obtain NIR spectra of all the PARSEC targets that currently lack NIR spectroscopy (see AHA11 for the complete target list).

The new parallaxes, proper motions and spectra presented here and in AHA11 will contribute to the creation of a large database of brown dwarfs. The creation of this database is one of the outputs of the Interpretation and Parameterization of Extremely Red COOL objects (IPERCOOL) International Research Staff Exchange Scheme, hosted on the IPERCOOL website (http://ipercool.oato.inaf.it).

\acknowledgments
This research is based on observations collected: at the European Organisation for Astronomical Research in the Southern Hemisphere, Chile programs 079.A-9203, 081.A-9200, 082.C-0946, 083.C-0446, 085.C-0690, 086.C-0168 and 186.C-0756; at the Southern Astrophysical Research (SOAR) telescope, which is a joint project of the Minist\'{e}rio da Ci\^{e}ncia, Tecnologia, e Inova\c{c}\~{a}o (MCTI) da Rep\'{u}blica Federativa do Brasil, the U.S. National Optical Astronomy Observatory (NOAO), the University of North Carolina at Chapel Hill (UNC), and Michigan State University (MSU). The SOAR/OSIRIS spectra were obtained as part of the proposals SO2009A-008, SO2011A-009 and SO2011B-006.

The authors would like to acknowledge the support of the Marie Curie 7th European Community Framework Programme grant n.236735 Parallaxes of Southern Extremely Cool objects (PARSEC) International Incoming Fellowship and grant n.247593 Interpretation and Parameterization of Extremely Red COOL dwarfs (IPERCOOL) International Research Staff Exchange Scheme. AHA thanks CNPq grant PQ-307126/2006-0. ADJ is supported by a FONDECYT postdoctorado fellowship under project number 3100098. ADJ is also partially supported by the Joint Committee ESO-Government Chile.

This research has made use of: the SIMBAD database operated at CDS France; the Two Micron All Sky Survey which is a joint project of the University of Massachusetts and the Infrared Processing and Analysis Center/California Institute of Technology; the SpeX Prism Spectral Libraries, maintained by Adam Burgasser at http://pono.ucsd.edu/$\sim$adam/browndwarfs/spexprism; and, the M, L, and T dwarf compendium housed at dwarfArchives.org and maintained by Chris Gelino, Davy Kirkpatrick, and Adam Burgasser.


\begin{thebibliography}{}

\bibitem[Allard et al.(2011)]{2011ASPC..448...91A} Allard, F., Homeier, D., \& Freytag, B.\ 2011, 16th Cambridge Workshop on Cool Stars, Stellar Systems, and the Sun, 448, 91 

\bibitem[Andrei et al.(2011)]{2011AJ....141...54A} Andrei, A.~H., Smart, R.~L., Penna, J.~L., et al.\ 2011, \aj, 141, 54 

\bibitem[Bakos et al.(2002)]{2002ApJS..141..187B} Bakos, G.~{\'A}., Sahu, K.~C., \& N{\'e}meth, P.\ 2002, \apjs, 141, 187 

\bibitem[Baraffe et al.(1998)]{1998A&A...337..403B} Baraffe, I., Chabrier, G., Allard, F., \& Hauschildt, P.~H.\ 1998, \aap, 337, 403 

\bibitem[Barman et al.(2011)]{2011ApJ...735L..39B} Barman, T.~S., Macintosh, B., Konopacky, Q.~M., \& Marois, C.\ 2011, \apjl, 735, L39 

\bibitem[Bensby et al.(2003)]{2003A&A...410..527B} Bensby, T., Feltzing, S., \& Lundstr{\"o}m, I.\ 2003, \aap, 410, 527 

\bibitem[Bouy et al.(2003)]{2003AJ....126.1526B} Bouy, H., Brandner, W., Mart{\'{\i}}n, E.~L., et al.\ 2003, \aj, 126, 1526 

\bibitem[Burgasser et al.(2006)]{2006ApJ...637.1067B} Burgasser, A.~J., Geballe, T.~R., Leggett, S.~K., Kirkpatrick, J.~D., \& Golimowski, D.~A.\ 2006, \apj, 637, 1067 

\bibitem[Burgasser et al.(2010)]{2010ApJ...710.1142B} Burgasser, A.~J., Cruz, K.~L., Cushing, M., et al.\ 2010, \apj, 710, 1142 

\bibitem[Burgasser(2013)]{2013AN....334...32B} Burgasser, A.~J.\ 2013, Astronomische Nachrichten, 334, 32 

\bibitem[Burningham et al.(2010)]{2010MNRAS.406.1885B} Burningham, B., Pinfield, D.~J., Lucas, P.~W., et al.\ 2010, \mnras, 406, 1885 

\bibitem[Burrows et al.(1997)]{1997ApJ...491..856B} Burrows, A., Marley, M., Hubbard, W.~B., et al.\ 1997, \apj, 491, 856 

\bibitem[Burrows et al.(2006)]{2006ApJ...640.1063B} Burrows, A., Sudarsky, D., \& Hubeny, I.\ 2006, \apj, 640, 1063 

\bibitem[Burrows et al.(2011)]{2011ApJ...736...47B} Burrows, A., Heng, K., \& Nampaisarn, T.\ 2011, \apj, 736, 47 

\bibitem[Chauvin et al.(2004)]{2004A&A...425L..29C} Chauvin, G., Lagrange, A.-M., Dumas, C., et al.\ 2004, \aap, 425, L29 

\bibitem[Clarke et al.(2010)]{2010MNRAS.402..575C} Clarke, J.~R.~A., Pinfield, D.~J., G{\'a}lvez-Ortiz, M.~C., et al.\ 2010, \mnras, 402, 575 

\bibitem[Cruz et al.(2003)]{2003AJ....126.2421C} Cruz, K.~L., Reid, I.~N., Liebert, J., Kirkpatrick, J.~D., \& Lowrance, P.~J.\ 2003, \aj, 126, 2421 

\bibitem[Cruz et al.(2007)]{2007AJ....133..439C} Cruz, K.~L., Reid, I.~N., Kirkpatrick, J.~D., et al.\ 2007, \aj, 133, 439 

\bibitem[Cruz et al.(2009)]{2009AJ....137.3345C} Cruz, K.~L., Kirkpatrick, J.~D., \& Burgasser, A.~J.\ 2009, \aj, 137, 3345 

\bibitem[Cushing et al.(2011)]{2011ApJ...743...50C} Cushing, M.~C., Kirkpatrick, J.~D., Gelino, C.~R., et al.\ 2011, \apj, 743, 50 

\bibitem[Day-Jones et al.(2013)]{2013MNRAS.430.1171D} Day-Jones, A.~C., Marocco, F., Pinfield, D.~J., et al.\ 2013, \mnras, 430, 1171 

\bibitem[Deacon \& Hambly(2007)]{2007A&A...468..163D} Deacon, N.~R., \& Hambly, N.~C.\ 2007, \aap, 468, 163 

\bibitem[Delfosse et al.(1999)]{1999A&AS..135...41D} Delfosse, X., Tinney, C.~G., Forveille, T., et al.\ 1999, \aaps, 135, 41 

\bibitem[Dupuy \& Liu(2011)]{2011ApJ...733..122D} Dupuy, T.~J., \& Liu, M.~C.\ 2011, \apj, 733, 122 

\bibitem[Dupuy \& Liu(2012)]{2012ApJS..201...19D} Dupuy, T.~J., \& Liu, M.~C.\ 2012, \apjs, 201, 19 

\bibitem[Epchtein et al.(1999)]{1999A&A...349..236E} Epchtein, N., Deul, E., Derriere, S., et al.\ 1999, \aap, 349, 236 

\bibitem[EROS Collaboration et al.(1999)]{1999A&A...351L...5E} EROS Collaboration, Goldman, B., Delfosse, X., et al.\ 1999, \aap, 351, L5 

\bibitem[Faherty et al.(2009)]{2009AJ....137....1F} Faherty, J.~K., Burgasser, A.~J., Cruz, K.~L., et al.\ 2009, \aj, 137, 1 

\bibitem[Faherty et al.(2012)]{2012ApJ...752...56F} Faherty, J.~K., Burgasser, A.~J., Walter, F.~M., et al.\ 2012, \apj, 752, 56 

\bibitem[Fan et al.(2000)]{2000AJ....119..928F} Fan, X., Knapp, G.~R., Strauss, M.~A., et al.\ 2000, \aj, 119, 928 

\bibitem[Finch et al.(2007)]{2007AJ....133.2898F} Finch, C.~T., Henry, T.~J., Subasavage, J.~P., Jao, W.-C., \& Hambly, N.~C.\ 2007, \aj, 133, 2898 

\bibitem[Gizis(2002)]{2002ApJ...575..484G} Gizis, J.~E.\ 2002, \apj, 575, 484 

\bibitem[Golimowski et al.(2004)]{2004AJ....127.3516G} Golimowski, D.~A., Leggett, S.~K., Marley, M.~S., et al.\ 2004, \aj, 127, 3516 

\bibitem[Hawley et al.(2002)]{2002AJ....123.3409H} Hawley, S.~L., Covey, K.~R., Knapp, G.~R., et al.\ 2002, \aj, 123, 3409 

\bibitem[Kendall et al.(2007)]{2007MNRAS.374..445K} Kendall, T.~R., Jones, H.~R.~A., Pinfield, D.~J., et al.\ 2007, \mnras, 374, 445 

\bibitem[Kirkpatrick et al.(1999)]{1999ApJ...519..802K} Kirkpatrick, J.~D., Reid, I.~N., Liebert, J., et al.\ 1999, \apj, 519, 802 

\bibitem[Kirkpatrick et al.(2000)]{2000AJ....120..447K} Kirkpatrick, J.~D., Reid, I.~N., Liebert, J., et al.\ 2000, \aj, 120, 447 

\bibitem[Kirkpatrick(2005)]{2005ARA&A..43..195K} Kirkpatrick, J.~D.\ 2005, \araa, 43, 195 

\bibitem[Kirkpatrick et al.(2006)]{2006ApJ...639.1120K} Kirkpatrick, J.~D., Barman, T.~S., Burgasser, A.~J., et al.\ 2006, \apj, 639, 1120 

\bibitem[Kirkpatrick et al.(2008)]{2008ApJ...689.1295K} Kirkpatrick, J.~D., Cruz, K.~L., Barman, T.~S., et al.\ 2008, \apj, 689, 1295 

\bibitem[Kirkpatrick et al.(2010)]{2010ApJS..190..100K} Kirkpatrick, J.~D., Looper, D.~L., Burgasser, A.~J., et al.\ 2010, \apjs, 190, 100 

\bibitem[Kirkpatrick et al.(2011)]{2011ApJS..197...19K} Kirkpatrick, J.~D., Cushing, M.~C., Gelino, C.~R., et al.\ 2011, \apjs, 197, 19 

\bibitem[Knapp et al.(2004)]{2004AJ....127.3553K} Knapp, G.~R., Leggett, S.~K., Fan, X., et al.\ 2004, \aj, 127, 3553 

\bibitem[Kurucz(1993)]{1993IAUCB..21...93K} Kurucz, R.~L.\ 1993, IAU Commission on Close Binary Stars, 21, 93 

\bibitem[Lawrence et al.(2007)]{2007MNRAS.379.1599L} Lawrence, A., Warren, S.~J., Almaini, O., et al.\ 2007, \mnras, 379, 1599 

\bibitem[Leggett et al.(2012)]{2012ApJ...748...74L} Leggett, S.~K., Saumon, D., Marley, M.~S., et al.\ 2012, \apj, 748, 74 

\bibitem[L{\'e}pine et al.(2007)]{2007ApJ...669.1235L} L{\'e}pine, S., Rich, R.~M., \& Shara, M.~M.\ 2007, \apj, 669, 1235 

\bibitem[Liebert et al.(2003)]{2003AJ....125..343L} Liebert, J., Kirkpatrick, J.~D., Cruz, K.~L., et al.\ 2003, \aj, 125, 343 

\bibitem[Lodieu et al.(2002)]{2002A&A...389L..20L} Lodieu, N., Scholz, R.-D., \& McCaughrean, M.~J.\ 2002, \aap, 389, L20 

\bibitem[Lodieu et al.(2005)]{2005A&A...440.1061L} Lodieu, N., Scholz, R.-D., McCaughrean, M.~J., et al.\ 2005, \aap, 440, 1061 

\bibitem[Looper et al.(2007)]{2007AJ....134.1162L} Looper, D.~L., Kirkpatrick, J.~D., \& Burgasser, A.~J.\ 2007, \aj, 134, 1162 

\bibitem[Looper et al.(2008)]{2008ApJ...685.1183L} Looper, D.~L., Gelino, C.~R., Burgasser, A.~J., \& Kirkpatrick, J.~D.\ 2008, \apj, 685, 1183 

\bibitem[Lucas et al.(2001)]{2001MNRAS.326..695L} Lucas, P.~W., Roche, P.~F., Allard, F., \& Hauschildt, P.~H.\ 2001, \mnras, 326, 695 

\bibitem[Marley et al.(2007)]{2007ApJ...655..541M} Marley, M.~S., Fortney, J.~J., Hubickyj, O., Bodenheimer, P., \& Lissauer, J.~J.\ 2007, \apj, 655, 541 

\bibitem[Marocco et al.(2010)]{2010A&A...524A..38M} Marocco, F., Smart, R.~L., Jones, H.~R.~A., et al.\ 2010, \aap, 524, A38 

\bibitem[Mart{\`i}n(2000)]{2000vlms.conf..119M} Mart{\`i}n, E.~L.\ 2000, Very Low-Mass Stars and Brown Dwarfs, 119 

\bibitem[Mart{\'{\i}}n et al.(2010)]{2010A&A...517A..53M} Mart{\'{\i}}n, E.~L., Phan-Bao, N., Bessell, M., et al.\ 2010, \aap, 517, A53 

\bibitem[Mart{\'{\i}}n et al.(2006)]{2006A&A...456..253M} Mart{\'{\i}}n, E.~L., Brandner, W., Bouy, H., et al.\ 2006, \aap, 456, 253 

\bibitem[Murray et al.(2011)]{2011MNRAS.414..575M} Murray, D.~N., Burningham, B., Jones, H.~R.~A., et al.\ 2011, \mnras, 414, 575 

\bibitem[Nielsen et al.(2012)]{2012ApJ...750...53N} Nielsen, E.~L., Liu, M.~C., Wahhaj, Z., et al.\ 2012, \apj, 750, 53 

\bibitem[Nissen(2004)]{2004oee..symp..154N} Nissen, P.~E.\ 2004, Origin and Evolution of the Elements, 154 

\bibitem[Patience et al.(2010)]{2010A&A...517A..76P} Patience, J., King, R.~R., de Rosa, R.~J., \& Marois, C.\ 2010, \aap, 517, A76 

\bibitem[Pinfield et al.(2006)]{2006MNRAS.368.1281P} Pinfield, D.~J., Jones, H.~R.~A., Lucas, P.~W., et al.\ 2006, \mnras, 368, 1281 

\bibitem[Pinfield et al.(2012)]{2012MNRAS.422.1922P} Pinfield, D.~J., Burningham, B., Lodieu, N., et al.\ 2012, \mnras, 422, 1922 

\bibitem[Reid(1992)]{1992MNRAS.257..257R} Reid, N.\ 1992, \mnras, 257, 257 

\bibitem[Reid et al.(2001)]{2001AJ....121.1710R} Reid, I.~N., Burgasser, A.~J., Cruz, K.~L., Kirkpatrick, J.~D., \& Gizis, J.~E.\ 2001, \aj, 121, 1710 

\bibitem[Reid et al.(2008)]{2008AJ....136.1290R} Reid, I.~N., Cruz, K.~L., Kirkpatrick, J.~D., et al.\ 2008, \aj, 136, 1290 

\bibitem[Rojas-Ayala et al.(2010)]{2010ApJ...720L.113R} Rojas-Ayala, B., Covey, K.~R., Muirhead, P.~S., \& Lloyd, J.~P.\ 2010, \apjl, 720, L113 

\bibitem[Saumon \& Marley(2008)]{2008ApJ...689.1327S} Saumon, D., \& Marley, M.~S.\ 2008, \apj, 689, 1327 

\bibitem[Schmidt et al.(2010)]{2010AJ....139.1808S} Schmidt, S.~J., West, A.~A., Hawley, S.~L., \& Pineda, J.~S.\ 2010, \aj, 139, 1808 

\bibitem[Scholz \& Meusinger(2002)]{2002MNRAS.336L..49S} Scholz, R.-D., \& Meusinger, H.\ 2002, \mnras, 336, L49 

\bibitem[Scholz et al.(2005)]{2005A&A...430L..49S} Scholz, R.-D., McCaughrean, M.~J., Zinnecker, H., \& Lodieu, N.\ 2005, \aap, 430, L49 

\bibitem[Seifahrt et al.(2005)]{2005AN....326..974S} Seifahrt, A., Mugrauer, M., Wiese, M., Neuh{\"a}user, R., \& Guenther, E.~W.\ 2005, Astronomische Nachrichten, 326, 974 

\bibitem[Skemer et al.(2011)]{2011ApJ...732..107S} Skemer, A.~J., Close, L.~M., Sz{\H u}cs, L., et al.\ 2011, \apj, 732, 107 

\bibitem[Skrutskie et al.(2006)]{2006AJ....131.1163S} Skrutskie, M.~F., Cutri, R.~M., Stiening, R., et al.\ 2006, \aj, 131, 1163 

\bibitem[Smart et al.(2003)]{2003A&A...404..317S} Smart, R.~L., Lattanzi, M.~G., Bucciarelli, B., et al.\ 2003, \aap, 404, 317

\bibitem[Smart et al.(2007)]{2007A&A...464..787S} Smart, R.~L., Lattanzi, M.~G., Jahrei{\ss}, H., Bucciarelli, B., \& Massone, G.\ 2007, \aap, 464, 787  

\bibitem[Stassun et al.(2006)]{2006Natur.440..311S} Stassun, K.~G., Mathieu, R.~D., \& Valenti, J.~A.\ 2006, \nat, 440, 311 

\bibitem[Stephens et al.(2009)]{2009ApJ...702..154S} Stephens, D.~C., Leggett, S.~K., Cushing, M.~C., et al.\ 2009, \apj, 702, 154 

\bibitem[Teixeira et al.(2008)]{2008A&A...489..825T} Teixeira, R., Ducourant, C., Chauvin, G., et al.\ 2008, \aap, 489, 825 

\bibitem[Tokunaga \& Kobayashi(1999)]{1999AJ....117.1010T} Tokunaga, A.~T., \& Kobayashi, N.\ 1999, \aj, 117, 1010 

\bibitem[Torres(1999)]{1999PASP..111..169T} Torres, G.\ 1999, \pasp, 111, 169 

\bibitem[Wright et al.(2010)]{2010AJ....140.1868W} Wright, E.~L., Eisenhardt, P.~R.~M., Mainzer, A.~K., et al.\ 2010, \aj, 140, 1868 

\bibitem[York et al.(2000)]{2000AJ....120.1579Y} York, D.~G., Adelman, J., Anderson, J.~E., Jr., et al.\ 2000, \aj, 120, 1579 

\end{thebibliography}
\end{document}